\documentclass[lettersize,journal]{IEEEtran}

\usepackage[normalem]{ulem}
\usepackage{xspace}
\usepackage{xcolor}
\usepackage[colorlinks=true,linkcolor=blue]{hyperref}
\usepackage[hypcap=false]{caption}
\usepackage{graphicx}
\usepackage{enumitem}
\usepackage{fancyvrb}
\usepackage{listings}
\usepackage{amsthm}
\usepackage{amsmath}
\usepackage{booktabs}
\usepackage{multirow}
\usepackage{bm}
\usepackage{textcomp}
\usepackage{tcolorbox}
\usepackage{colortbl}
\usepackage{makecell}
\usepackage{url}
\usepackage[ruled,vlined,linesnumbered]{algorithm2e}
\usepackage[switch]{lineno} 
\usepackage{algpseudocode}
\usepackage{bbding}

\newcommand{\Toolname}{\textsc{HyLLfuzz}}
\newcommand{\tool}{{\sc \Toolname}\xspace}
\newcommand{\toolBold}{{\sc\bfseries \Toolname}\xspace}
\newcommand{\Ablationname}{\textsc{HF0}}
\newcommand{\ablation}{{\sc \Ablationname}\xspace}
\newcommand{\driller}{{\sc Driller}\xspace}
\newcommand{\cofuzz}{{\sc CoFuzz}\xspace}
\newcommand{\cofuzzBold}{{\sc\bfseries CoFuzz}\xspace}
\newcommand{\qsym}{{\sc QSYM}\xspace}
\newcommand{\qsymBold}{{\sc\bfseries QSYM}\xspace}
\newcommand{\intriguer}{{\sc Intriguer}\xspace}
\newcommand{\intriguerBold}{{\sc\bfseries Intriguer}\xspace}
\newcommand{\digfuzz}{{\sc DigFuzz}\xspace}
\newcommand{\meuzz}{{\sc MEUZZ}\xspace}
\newcommand{\pangolin}{{\sc Pangolin}\xspace}
\newcommand{\llmdfa}{{\sc LLMDFA}\xspace}
\newcommand{\llift}{{\sc LLift}\xspace}
\newcommand{\codamosa}{{\sc CodaMOSA}\xspace}
\newcommand{\afl}{{\sc AFL}\xspace}
\newcommand{\klee}{{\sc KLEE}\xspace}
\newcommand{\nautilus}{{\sc Nautilus}\xspace}
\newcommand{\nautilusBold}{{\sc\bfseries Nautilus}\xspace}
\newcommand{\superion}{{\sc Superion}\xspace}
\newcommand{\superionBold}{{\sc\bfseries Superion}\xspace}
\newcommand{\langfuzz}{{\sc LangFuzz}\xspace}

\newcommand{\jazzer}{{\sc Jazzer}\xspace}

\newcommand{\chatafl}{{\sc ChatAFL}\xspace}
\newcommand{\fuzzbench}{{\sc FuzzBench}\xspace}
\newcommand{\magma}{{\sc Magma}\xspace}
\newcommand{\unifuzz}{{\sc UniFuzz}\xspace}
\newcommand{\dart}{{\sc DART}\xspace}
\newcommand{\cute}{{\sc CUTE}\xspace}
\newcommand{\exe}{{\sc EXE}\xspace}
\newcommand{\ste}{{\sc S2E}\xspace}
\newcommand{\coverup}{{\sc CoverUp}\xspace}
\newcommand{\testpilot}{{\sc TestPilot}\xspace}
\newcommand{\chatunittest}{{\sc ChatUniTest}\xspace}
\newcommand{\symprompt}{{\sc SymPrompt}\xspace}
\newcommand{\fuzzall}{{\sc Fuzz4All}\xspace}

\newcommand{\llmfuzz}{{\sc LLM4Fuzz}\xspace}
\newcommand{\titanfuzz}{{\sc TitanFuzz}\xspace}
\newcommand{\hits}{{\sc HiTS}\xspace}

\definecolor{deepblue}{rgb}{0,0,0.5}
\definecolor{deepred}{rgb}{0.6,0,0}
\definecolor{deepgreen}{rgb}{0,0.5,0}
\definecolor{halfgray}{gray}{0.55}
\definecolor{ipythonframe}{RGB}{207, 207, 207}
\definecolor[named]{ACMBlue}{cmyk}{1,0.1,0,0.1}
\definecolor[named]{ACMYellow}{cmyk}{0,0.16,1,0}
\definecolor[named]{ACMOrange}{cmyk}{0,0.42,1,0.01}
\definecolor[named]{ACMRed}{cmyk}{0,0.90,0.86,0}
\definecolor[named]{ACMLightBlue}{cmyk}{0.49,0.01,0,0}
\definecolor[named]{ACMGreen}{cmyk}{0.20,0,1,0.19}
\definecolor[named]{ACMPurple}{cmyk}{0.55,1,0,0.15}
\definecolor[named]{ACMDarkBlue}{cmyk}{1,0.58,0,0.21}

\newcommand{\ignore}[1]{}

\definecolor{shadecolor}{gray}{1.00}
\definecolor{ddarkgray}{gray}{0.75}
\definecolor{darkgray}{gray}{0.30}
\definecolor{light-gray}{gray}{0.87}

\hypersetup{colorlinks,
  linkcolor=ACMDarkBlue,
  citecolor=ACMPurple,
  urlcolor=ACMDarkBlue,
  filecolor=ACMDarkBlue}

\newcommand{\ie}{i.e.,\xspace}

\newcommand{\eg}{e.g.,\xspace}

\newcommand{\ccc}[1]{\raisebox{.9pt}{\textcircled{\raisebox{-.9pt}{\small #1}}}}

\theoremstyle{definition}

\definecolor{mycolor}{rgb}{0.122, 0.435, 0.698}
\newcommand{\result}[1]{%
\begin{tcolorbox}[colframe=mycolor,boxrule=0.5pt,arc=4pt,
      left=6pt,right=6pt,top=6pt,bottom=6pt,boxsep=0pt,width=\columnwidth]%
      {#1}
\end{tcolorbox}%
}

\definecolor{keyword}{rgb}{0.0, 0, 0.5}
\definecolor{comment}{rgb}{0.0, 0.5, 0.0}
\definecolor{const}{rgb}{0.6,0,0}
\definecolor{customorange}{rgb}{0.8863, 0.6118, 0.2706}
\definecolor{syscall}{rgb}{0.4,0.4,0}
\definecolor{stoneblue}{rgb}{0.086, 0.522, 0.663}
\definecolor{customgreen}{rgb}{0.482, 0.812, 0.651}
\definecolor{background}{rgb}{0.95,0.95,0.95}

\lstdefinelanguage{MyC}{
    language=C,
    keywordstyle=\color{keyword}\ttfamily\fontseries{b}\selectfont,
    keywordstyle=[2]\color{const}\ttfamily\fontseries{b}\selectfont,
    keywordstyle=[3]\color{syscall}\ttfamily\fontseries{b}\selectfont,
    morekeywords=[2]{SIZE},
    morekeywords=[3]{read,write,recv,send,exit},
    literate=%
    {printf}{{\textcolor{customorange}{printf}}}6
    {free}{{\textcolor{stoneblue}{free}}}4
    {target}{{\textcolor{const}{target}}}5
    {memcmp}{{\textcolor{customorange}{memcmp}}}5
    {'}{{\textquotesingle}}1
}

\lstdefinelanguage{MyJava}{
    language=Java,
    keywordstyle=\color{keyword}\ttfamily\fontseries{b}\selectfont,
    keywordstyle=[2]\color{const}\ttfamily\fontseries{b}\selectfont,
    keywordstyle=[3]\color{syscall}\ttfamily\fontseries{b}\selectfont,
    morekeywords=[2]{"x-evil-backdoor"},
    morekeywords={System.out.println, new, public, static, void, main, int, float, double, String},
    literate=%
    {vulnerable}{{\textcolor{const}{vulnerable}}}8
}

\usepackage[normalem]{ulem} 
\newcommand\hl{\bgroup\markoverwith
  {\textcolor{gray!30}{\rule[-.5ex]{2pt}{2.5ex}}}\ULon}

\def\BibTeX{{\rm B\kern-.05em{\sc i\kern-.025em b}\kern-.08em
    T\kern-.1667em\lower.7ex\hbox{E}\kern-.125emX}}

\begin{document}


\title{Large Language Model assisted Hybrid Fuzzing}

\author{
    Ruijie Meng,
    Gregory J. Duck,
    and Abhik Roychoudhury%
    \IEEEcompsocitemizethanks{
        \IEEEcompsocthanksitem
        R. Meng is with the National University of Singapore, Singapore.
        E-mail: ruijie\_meng@u.nus.edu.
        \IEEEcompsocthanksitem
        G. Duck is with the National University of Singapore, Singapore.
        E-mail: gregory@comp.nus.edu.sg.
        \IEEEcompsocthanksitem
        A. Roychoudhury is with the National University of Singapore, Singapore.
        E-mail: abhik@nus.edu.sg.
    }
}

\maketitle

\begin{abstract}
Greybox fuzzing is one of the most popular methods for detecting software vulnerabilities, which conducts a biased random search within the program input space. To enhance its effectiveness in achieving deep coverage of program behaviors, greybox fuzzing is often combined with concolic execution, which performs a path-sensitive search over the domain of program inputs. In hybrid fuzzing, conventional greybox fuzzing is followed by concolic execution in an iterative loop, where reachability roadblocks encountered by greybox fuzzing are tackled by concolic execution. However, such hybrid fuzzing still suffers from difficulties conventionally faced by concolic execution, such as the need for environment modeling and system call support. In this work, we explore the potential of developing ``smart'' concolic execution empowered by Large Language Models (LLMs), leveraging their knowledge of code semantics during constraint computing and solving. When coverage-based greybox fuzzing reaches a roadblock in terms of reaching certain branches, we conduct a slicing on the execution trace and suggest modifications of the input to reach the relevant branches. The LLM is used as a solver to generate the modified input to reach the desired branches. Compared with state-of-the-art hybrid fuzzers \cofuzzBold, \intriguerBold, and \qsymBold, our LLM-based hybrid fuzzer \toolBold (pronounced ``{\em hill fuzz}'') covers 31.43\%, 44.56\%, and 59.48\% more code branches, respectively. Furthermore, the LLM-based concolic execution in \toolBold takes a time that is 3--19 times faster than the concolic execution running in existing hybrid fuzzing tools. In extensively tested real-world subjects, \toolBold exposed seven previously unknown bugs. This experience shows that LLMs can be effectively inserted into the iterative loop of hybrid fuzzers to efficiently expose more program behaviors.
\end{abstract}


\section{Introduction} \label{sec:introduction}

Greybox fuzzing \cite{fuzzing, aflfast} is one of the most widely used techniques today to detect security vulnerabilities in software systems. At its core, greybox fuzzing performs an adaptive and biased random search over the input space to find crashing inputs. It conducts lightweight compile-time instrumentation on selected program elements (e.g., branches), and at runtime records which instrumented elements are exercised for a given input. This runtime feedback then guides the input mutation to generate new ones and explore more program behaviors. 

Although greybox fuzzing is highly effective, it often gets stuck at complex branches that are rarely evaluated as true, such as those guarded by magic bytes (e.g., \verb+if(input == "\uDB16\uDC06")+). In such cases, concolic execution is more effective in generating inputs to cover the branches. Concolic execution \cite{dart, klee, symcc} searches the domain of inputs like greybox fuzzing, except that navigation is achieved by mutating path constraints instead of mutating inputs. A concolic execution engine executes the program with a concrete input while symbolically emulating the program along the same execution path to compute path constraints. By negating these path constraints, it transitions to alternative branches. A constraint solver is then used to solve the negated constraints and generate concrete inputs that drive the execution along new paths. However, its effectiveness often comes at the cost of efficiency \cite{bohme2020fuzzing, hybrid}.

Hybrid fuzzing \cite{hybrid} combines greybox fuzzing with concolic execution to leverage the strengths of both techniques and maximize code coverage. It was first introduced in \driller \cite{driller} and has subsequently been followed by many testing engines \cite{qsym, intriguer, digfuzz}. In hybrid fuzzing, greybox fuzzing proceeds until it encounters roadblocks in unraveling a branch, which is evaluated to be almost always true or false. This roadblock branch captures the frontier of the greybox fuzzing search. Concolic execution is then used to overcome this frontier. Once the roadblock is overcome, greybox fuzzing can take over again until another roadblock is encountered. 

Unfortunately, hybrid fuzzing remains limited due to the underlying concolic execution. Traditional concolic execution \emph{mechanically} computes constraints and solves them for satisfiability, without knowledge of the program semantics. This leads to several challenges. For example, real-world constructs, such as external libraries and system calls, require explicit modeling to allow further analysis, which may require years of manual effort \cite{s2e, meng2024program, pandey2019deferred}. Moreover, even when constraints are satisfiable, traditional solvers (e.g., \verb|Z3| \cite{de2008z3}) cannot prioritize solutions that also conform to the required input structure, potentially generating a number of syntactically invalid inputs. Compounding the problem, program paths in programming languages (e.g. \verb|C/C++| and \verb+Java+) must be symbolically translated as logic constraints that are solvable for solvers. However, both constraint computing and constraint solving are time-consuming \cite{qsym, intriguer, symcc}. 

Recent advances in Large Language Models (LLMs) open up opportunities for developing \emph{smart} concolic execution to address these challenges. Similar to traditional constraint solvers, LLMs have shown the capability of analyzing execution paths and then generating inputs \cite{chatafl, llm4fuzz}. However, unlike traditional solvers, LLMs have a knowledge of the program semantics. LLMs trained on terabytes of data corpus can understand external libraries \cite{jiang2024towards}, eliminating the need for manual modeling, and can infer code functionalities \cite{hou2023large, symprompt}, allowing the generation of syntactically valid inputs. Furthermore, LLMs are more adaptive and capable of directly reasoning in high-level programming languages. This allows the LLM-based approach to bypass the translation of source code to logic formulas, thus accelerating the analysis process. 

Building on this, we propose \tool, a hybrid fuzzer that integrates LLM-based concolic execution. Specifically, when greybox fuzzing reaches a coverage plateau, where it has difficulty covering new program paths, \tool identifies the roadblocks to overcome. The subsequent steps align with the principles of traditional concolic execution. \tool executes the input that reaches the branch condition, but fails to satisfy it. Meanwhile, \tool constructs a dynamic slice of the relevant execution path. Instead of translating the execution path into logic formulas, \tool preserves the path as-is in the form of a source code slice. The LLM is then prompted to modify the input based on the sliced code path and generate new inputs to overcome roadblocks. These LLM-generated inputs are added to the seed corpus, where they are retrieved and evaluated by greybox fuzzing. Whenever greybox fuzzing hits a coverage plateau again, this process is repeated. 

The embodiment of our approach in the form of \tool, shows strong promise in delivering \emph{effectiveness}, \emph{efficiency}, and {\em usability}. To the best of our knowledge, previous hybrid fuzzers have not considered all three of these important dimensions. In terms of effectiveness, \tool achieves 31--59\% more branch coverage on average compared to state-of-the-art hybrid fuzzers. In terms of efficiency (median time), each concolic execution run of \tool consumes 3--19 times less time compared to those of state-of-the-art hybrid fuzzers. Finally, in terms of usability, \tool by definition does not involve any symbolic modeling of instructions or building hand-crafted environment models, thereby naturally lowering the barrier to entry. 

In summary, we make the following main contributions:
\begin{itemize}[leftmargin=*]
    \item We propose an LLM-based concolic execution engine, with knowledge of the code semantics during input generation. This approach addresses the problems that affect traditional concolic execution engines.

    \item We implemented our approach in a novel hybrid fuzzer \tool, which combines greybox fuzzing with LLM-based concolic execution invoked whenever a roadblock is encountered. Our tool is publicly available at: 
    \begin{center}
        \url{https://github.com/mengrj/hyllfuzz}
    \end{center}

    \item We evaluated \tool on several widely used benchmarks, and compared it with state-of-the-art hybrid fuzzers. The results demonstrate that \tool delivers significantly greater effectiveness, efficiency, and usability.
\end{itemize}
\section{Motivating Example} \label{sec:motivation}

To motivate our approach, we use the \texttt{libxml2} example to demonstrate how LLM-based concolic execution is integrated into hybrid fuzzing to address the challenges. \texttt{libxml2} \cite{libxml2} is a widely used \texttt{XML} parser and also a common subject for evaluating fuzzing performance \cite{fuzzbench, Magma, unifuzz}. The code from \texttt{parser.c} in \autoref{fig:bkg_code} shows how \texttt{libxml2} parses an \texttt{XML} comment. The magic values (line 25) and nested conditions (lines 10--12) are roadblocks for greybox fuzzing, and hybrid fuzzing thus invokes concolic execution to solve them.  

{
  \begin{figure}[t]
  \footnotesize
  \begin{center}
  \begin{minipage}{\linewidth}
  \begin{lstlisting}[language=MyC, frame=single, numbers=left, breaklines=true,
  numbersep=5pt,           
  xleftmargin=15pt,        
  framexleftmargin=12pt,
  numberstyle=\color{black}\ttfamily,
  mathescape=true
  ]
/**
 * Parse an XML (SGML) comment
 * Comment::='<!--'((Ch-'-')|('-'(Ch-'-')))*'-->'
 */
void xmlParseComment(xmlParserCtxtPtr ctxt){
  ...
  do{
    // Skip 20+ path conditions
    printf("%c", *in);
    if (*in == '-') {
      if (in[1] == '-'){
        if (in[2] == '>'){
          return;
        } in++; 
      } goto get_more;
    } 
  } while (((*in>=0x20) && (*in<=0x7F)) || (*in==0x09) || (*in==0x0a));
  ...
}

void xmlParseContentInternal(xmlParserCtxtPtr ctxt){
  const xmlchatr *cur = ctxt->input->cur;
  ...
  /* This case: a comment */
  if ((*cur == '<') && (NXT(1) == '!') && (NXT(2) == '-') && (NXT(3) == '-')){
    xmlParseComment(ctxt);
  }
  ...
}
\end{lstlisting}
\end{minipage}
\end{center}
\caption{Code fragment extracted from \texttt{parser.c} in \texttt{libxml2}.}
\label{fig:bkg_code}
\end{figure}
}

Let us first take a look at how traditional hybrid fuzzing works. To apply concolic execution to the real-world code, traditional engines must abstract any system calls and external libraries (e.g., \verb|printf| in \autoref{fig:bkg_code}) behind models, which allows symbolic reasoning over execution paths. These models are typically handwritten. For example, \klee \cite{klee} includes over 2500 lines of \verb|C| code to support only 40 system calls, although the Linux kernel has more than 300 system calls. Writing these models is usually labor-intensive, even taking multiple person-years in some cases \cite{s2e}. 

During fuzzing, greybox fuzzing generates inputs that reach complex branch conditions (e.g., line 25 in \autoref{fig:bkg_code}) but fail to satisfy them, as they are always evaluated to false. In such cases, concolic execution is invoked. Starting from an input generated by greybox fuzzing, concolic execution executes it while emulating the code path to compute symbolic constraints, such as \verb|(= (select input i) 60)| for \texttt{(*cur == \textquotesingle<\textquotesingle)}. However, symbolic emulation can introduce significant time overhead, such as being several orders of magnitude slower than native execution \cite{symcc, intriguer, qsym}. The condition on line 25 checks if the string represents the opening delimiter (\verb|<!--|) of an \texttt{XML} comment, which involves four separate path constraints. Since concolic execution can only negate and solve one constraint per round, covering the full branch requires multiple rounds. Furthermore, a single round of constraint solving may take up to 500 seconds \cite{intriguer}. 

{
\begin{figure*}[t]
  \centering
  \includegraphics[page=1, trim= 0.28in 9.7in 1.3in 0.3in, clip, scale=1.06]{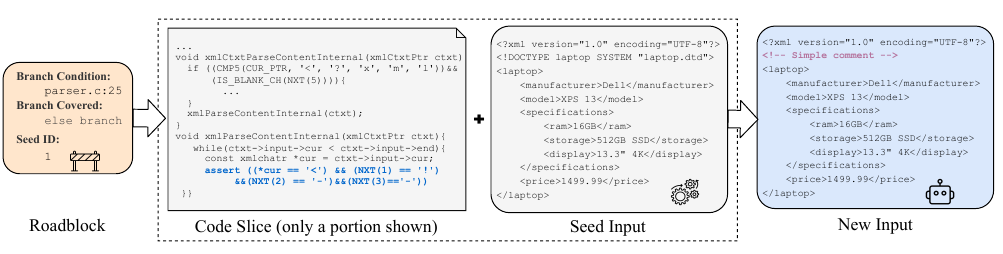}
  \caption{Workflow of solving the roadblock at \texttt{parser.c:25} based on a seed input (selected from the seed corpus of the greybox fuzzer) and the constraints related to the roadblock (sliced from source code) to generate a new input by the LLM.}
  \label{fig:motivation}
\end{figure*}
}

In addition, constraint solvers solve constraints just for satisfiability, without considering syntactic validity. When solving the constraints for line 17, solvers prioritize the value \verb|0x20|, thereby generating syntactically invalid inputs that lack the closing delimiter (\verb|-->|). Such invalid inputs are immediately rejected by the parser and thus fail to cover deeper program behaviors. Although the additional path conditions (lines 10--12) can eventually guide the solver toward generating valid inputs, doing so requires satisfying more than 20 path conditions (as shown in \autoref{fig:bkg_code}). This further reduces the efficiency of concolic execution. In our experiments, traditional concolic execution failed to generate a syntactically valid \texttt{XML} comment for input. In the best case, the engines only generated inputs with a partial opening delimiter (\eg \verb|<!|), which are unable to explore subsequent program behaviors. 

In this paper, we propose a novel hybrid fuzzer \tool, which integrates LLM-based concolic execution to address the above problems. The workflow of \tool to solve the roadblock in \texttt{parser.c:25} is shown in \autoref{fig:motivation}. \tool relies on LLMs to directly reason about the underlying functionalities of external constructs, without requiring manual modeling. During fuzzing, when the greybox fuzzer reaches a coverage plateau, \tool activates its LLM-based concolic execution. It first identifies the roadblock condition (\texttt{parser.c:25}) and determines the already covered branch (\texttt{else branch}). It then finds the corresponding input that reaches this condition, denoted \texttt{ID:1} by the greybox fuzzer. All of this information forms \emph{Roadblock}. 

\tool next performs a concrete execution using the selected input and collects the corresponding execution trace. It performs backward code slicing along the trace. Specifically, it expresses the target condition as an assertion and then slices the relevant code that affects the variables in this assertion (i.e., \texttt{cur} and \texttt{NXT}) through the control and data flow. This \emph{code slice} represents the path constraints. Since LLMs can directly reason over source code written in programming languages, \tool just keeps them as-is, and there is no need for constraint translation. Using the code slice and the seed input, the LLM is prompted to generate a \emph{new input} to satisfy the constraints, which is then evaluated by the greybox fuzzer. In our experiments, for a single execution, our code slicing combined with LLM-based solving is several times faster than traditional constraint computing and solving. 

\tool treats the entire condition, including four constraints, as a single solving target. Furthermore, from context information, such as function names (\eg \texttt{xmlParseComment}) and code comments (\eg \texttt{Parse an XML (SGML) comment}), LLMs infer that the code is parsing the opening delimiter of an \texttt{XML} comment. It then automatically appends the closing delimiter, generating a syntactically valid input that naturally covers subsequent branches (\eg lines 10--12). In contrast, traditional solvers are limited to modifying only symbolized data fields. This capability further enhances the effectiveness of \tool. In \texttt{libxml2}, \tool can cover twice as many code branches as state-of-the-art hybrid fuzzers. 
\section{LLM-assisted Hybrid Fuzzing} \label{sec:approach}

\subsection{Overview}

{
\begin{figure*}[t]
  \centering
  \includegraphics[page=1, trim= 0.4in 9.85in 0.25in 0.4in, clip, scale=0.88]{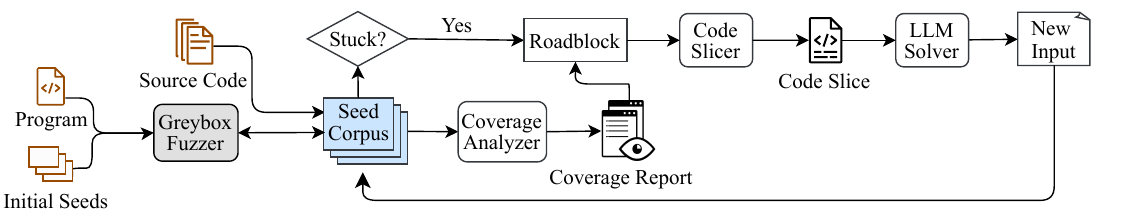}
  \caption{Overall workflow of \tool including greybox fuzzing and LLM-based concolic execution.}
  \label{fig:overview}
\end{figure*}
}

\tool is a hybrid fuzzer that combines greybox fuzzing with LLM-based concolic execution to maximize code coverage. It is designed as a plug-in that can seamlessly integrate with greybox fuzzers, without requiring any modifications to their source code. The overall workflow of \tool is illustrated in \autoref{fig:overview}. The process is initiated by the greybox fuzzer, which takes in as input the program under test and initial input seeds provided by the user. The greybox fuzzer iteratively selects inputs, generates new inputs via mutation, and saves ``interesting'' inputs (that cover new branches) into the seed corpus for further examination. 

Running in parallel, \tool continuously analyzes the code covered by the seed corpus and maintains a report of uncovered branches (\autoref{sec:coverage_report}). When the greybox fuzzer struggles to cover new branches within a given time, \tool steps in by identifying roadblocks from the maintained coverage report. Each roadblock consists of a partially covered branch condition, the branches already covered for this condition, and an input from the seed corpus that reaches this condition. \tool then simulates traditional concolic execution by running the program with the selected input and generating the relevant path slice from the source code (\autoref{sec:constraint_collect}). \tool negates the target branch condition and prompts the LLM to generate a new input that covers the previously uncovered branch (\autoref{sec:test_generate}). The newly generated input is added back into the seed corpus, assisting the greybox fuzzer in overcoming the roadblocks. The seed corpus forms a bridge between the greybox fuzzer and the LLM-based concolic execution engine.

\subsection{Code Coverage Analysis} \label{sec:coverage_report}

\subsubsection{{Generating Coverage Report}} The first step of \tool is to analyze branch coverage using inputs from the seed corpus generated by the greybox fuzzer. This aims to identify the branch conditions that (1) are reached by at least one input, and (2) have at least one branch not covered yet by the greybox fuzzer. Such conditions become potential targets for LLM-based concolic execution. \tool does not rely on the coverage bitmap recorded by the greybox fuzzer, as this bitmap only provides an approximate measure of branch coverage and is known to suffer from collisions~\cite{collafl}. Instead, \tool maintains an independent branch coverage report by directly analyzing inputs stored in the seed corpus. This process is illustrated in \autoref{alg:cover_analysis}. 

\tool works at the source code level. It first builds the Abstract Syntax Tree (AST) for the program (line 3). Next, \tool retrieves the inputs from the seed corpus $\mathit{S}$ of the greybox fuzzer and executes each to collect the corresponding execution trace (line 5). Here, the trace is represented as a sequence of code statements arranged in execution order, where each statement is labeled by its location in source code. 

For each statement in the trace, \tool checks if it is a conditional statement (e.g., \texttt{if}- or \texttt{switch-} statement) (line 7). If so, \tool further determines which branches of this condition are covered by the trace (line 8). Simultaneously, \tool computes the total branch number for this condition, which is used to determine whether all branches are covered by the full inputs from $S$. This information is then recorded in the coverage report $\mathcal{R}$ (line 9). \tool also tracks the corresponding seed in $\mathcal{R}$ for concolic execution. Furthermore, \tool maintains partially covered conditions in a separate structure $\mathcal{I}$ (lines 10--12), which supports the selection of roadblocks in the next phase. It is worth noting that, similar to traditional concolic execution, \tool only tracks conditions that are actually reached by seeds. Any condition that has not been reached yet is not a target for any concolic execution.

\begin{algorithm}[t]
    \caption{Analysis of Coverage and Roadblocks\label{alg:cover_analysis}}
    \DontPrintSemicolon
    \small
    \newcommand\mycommfont[1]{\ttfamily\textcolor{blue}{#1}}
    \SetCommentSty{mycommfont}
    \SetNoFillComment 
    \SetKw{Continue}{continue}
    \SetKwProg{Fn}{func}{:}{}
    \SetKwInOut{Input}{Input}
    \SetKwInOut{Output}{Output}
    \SetKwInOut{Globals}{Globals}
    \SetKwFunction{coverage}{AnalyzeCodeCoverage}
    \SetKwFunction{execute}{execute}
    \SetKwFunction{remove}{remove}
    \SetKwFunction{buildast}{BuildAST}
    \SetKwFunction{iscond}{isCond}
    \SetKwFunction{iscovered}{isFullyCovered}
    \SetKwFunction{pop}{pop}
    \SetKwFunction{maxvalue}{maxValue}
    \SetKwFunction{reduceinterest}{DecreasePriority}
    \SetKwFunction{analyzebranch}{AnalyzeBrn}
    \SetKwFunction{getbranch}{GetBrnCovered}
    \SetKwFunction{getseed}{GetSeedForBrn}
    \SetKwFunction{getroadblock}{RetrieveRoadblock}
    \SetKwFunction{roadblock}{Roadblock}
    \Input{Program $\mathcal{P}$, source code $\mathcal{C}$, seed corpus $\mathcal{S}$}
    \Globals{Coverage report $\mathcal{R}$, partially covered conditions $\mathcal{I}$}
    \Output{Roadblock $\mathcal{B}$ to be solved by concolic execution}
    \vspace{0.5em}

    $\mathcal{R} \leftarrow \emptyset$; $\mathcal{I} \leftarrow \emptyset$ \\
    \Fn{\coverage{$\mathcal{P}$, $\mathcal{C}$, $\mathcal{S}$}}{
        $\mathit{ast} \leftarrow \buildast{$\mathcal{C}$}$ \\
        \For{$\mathit{seed} \in \mathcal{S}$}{
            $\mathit{trace} \leftarrow \execute(\mathcal{P}, \mathit{seed})$ \\
            \ForEach{$\mathit{stmt} \in \mathit{trace}$}{
                \lIf{$\neg\iscond{$\mathit{stmt}$, $\mathit{ast}$}$}{\Continue}
                    $\mathit{branchCovd} \leftarrow \analyzebranch{$\mathit{stmt}$, $\mathit{ast}$, $\mathit{trace}$}$ \\
                    $\mathcal{R} \leftarrow \mathcal{R} \cup \{\langle \mathit{stmt}, \mathit{branchCovd}, \mathit{seed} \rangle\}$ \\
                   \If{\iscovered{$\mathit{stmt}$, $\mathcal{R}$}}{
                        $\mathcal{I} \leftarrow \mathcal{I} \backslash \{\mathit{stmt}\}$
                    }
                   \lElseIf {$\mathit{stmt} \notin \mathcal{I}$}{
                        $\mathcal{I} \leftarrow \mathcal{I} \cup \{\mathit{stmt}\}$
                    }
            }
        }
    }

    \Fn{\getroadblock{$\mathcal{P}$, $\mathcal{C}$, $\mathcal{S}$}}{
        $\coverage{$\mathcal{P}$, $\mathcal{C}$, $\mathcal{S}$}$ \\
        $\mathit{condStmt} \leftarrow \pop{$\mathcal{I}$}$ \\
        $\mathit{branchCovd} \leftarrow \getbranch{$\mathcal{R}$, $\mathit{condStmt}$}$ \\
        $\mathit{seed} \leftarrow \getseed{$\mathcal{R}$, $\mathit{condStmt}$, $\mathit{branchCovd}$}$ \\
        $\mathcal{I} \leftarrow \reduceinterest{$\mathcal{I}$, $\mathit{condStmt}$}$ \\
        \Return $\mathcal{B}$$\langle \mathit{condStmt}, \mathit{branchCovd}, \mathit{seed} \rangle$
    }
\end{algorithm}

\subsubsection{{Selecting a Roadblock}} \tool continuously monitors whether the greybox fuzzer fails to cover new branches within the given time. To do this, \tool checks the time elapsed since the greybox fuzzer found the last new code path. If the duration exceeds the allotted threshold, the greybox fuzzer is considered to have entered a coverage plateau. At this point, \tool iteratively performs a code coverage analysis and identifies roadblocks from the coverage report $\mathcal{R}$ and partially covered conditions $\mathcal{I}$. 

To ensure fair selection, \tool arranges $\mathcal{I}$ as \emph{priority queue}, assigning the highest priority to newly added conditions. Once a condition is selected, its priority is reduced to avoid starvation. However, if solving a condition successfully generates a new input that increases the code coverage, as validated by the greybox fuzzer (illustrated in \autoref{sec:test_generate}), its priority is increased. This helps prioritize conditions that result in more coverage. Finally, once all branches of a condition are covered, it is removed from $\mathcal{I}$, since it no longer contains roadblock branches (line 11).

After selecting a partially covered condition $\mathit{condStmt}$ from $\mathcal{I}$ (line 15), \tool determines which of its branches has been covered ($\mathit{branchCovd}$) (line 16). For example, for an \texttt{if}-statement, it indicates which branch, \texttt{then-branch} or \texttt{else-branch}, is covered. \tool then invokes concolic execution by negating the branch condition to explore previously uncovered branches. It also retrieves the corresponding input for the concrete execution (line 17). These three elements together define a {\em roadblock}: (1) a partially covered condition ($\mathit{condStmt}$), (2) the branches covered ($\mathit{branchCovd}$), and (3) an input ($\mathit{seed}$) that reaches this condition (line 19).

\subsection{Code Slice Generation} \label{sec:constraint_collect}

\begin{algorithm}[t]
\caption{Dynamic Code Slice Generation}
\label{alg:slicing}
\small
\KwIn{Program $\mathcal{P}$, source code $\mathcal{C}$, roadblock $\mathcal{B}$}
\KwOut{Dynamic code slice (path constraints)}

\newcommand\mycommfont[1]{\footnotesize\ttfamily\textcolor{blue}{#1}}
\SetCommentSty{mycommfont}
\SetNoFillComment
\SetKwProg{Fn}{func}{:}{}
\SetKwFunction{execute}{execute}
\SetKwFunction{buildSlice}{BuildSlice}
\SetKwFunction{negateCoverage}{NegateCoverage}
\SetKwFunction{getVars}{getVars}
\SetKwFunction{getDependencies}{getDependencies}
\SetKwFunction{getContext}{getCtx}
\SetKwFunction{flattenPathConstraint}{FlattenSlice}
\SetKwFunction{buildast}{BuildAST}

\DontPrintSemicolon
\vspace{0.5em}

\Fn{\buildSlice{$\mathcal{P}$, $\mathcal{C}$, $\mathcal{B}$}}{
    \label{line:slicer_begin}
    $\langle \mathit{condStmt}, \mathit{branchCovd}, \mathit{seed} \rangle \leftarrow \mathcal{B}$ \\
    $\mathit{ast} \leftarrow \buildast{$\mathcal{C}$}$ \\
    $\langle s_1, ..., s_n, \mathit{condStmt}, ... \rangle \leftarrow \execute{$\mathcal{P}$, $\mathit{seed}$}$ \\
    $\mathit{stmts} \gets \{\texttt{assert(!}\mathit{branchCovd}\texttt{)})\}$ \\
    $\mathit{vars} \gets \getVars(\mathit{condStmt})$\\
    \For{$i \in n .. 1$}{
        \If{$s_i$ modifies any variable $v \in \mathit{vars}$, \textbf{or} \\
            \hspace{.7em} any condition from $s_i$ reads any $v \in \mathit{vars}$ \\
            }{
            $\mathit{stmts} \gets \mathit{stmts} \cup \{s_i\}$\\
            $\mathit{vars} \gets \mathit{vars} \cup \getDependencies(v, s_i)$\label{line:slicer_end}\;
        }
    }
    $\mathit{slice} \gets \emptyset$\label{line:render_begin}\;
    \For{$s \in \mathit{stmts}$}{
       $\mathit{slice} \gets \mathit{slice} \cup \{s\} \cup \getContext(\mathit{ast}, s)$\\
    }
    \Return{$\flattenPathConstraint(\mathit{slice})$}\label{line:render_end}\;
}

\end{algorithm}

Once a roadblock is selected, the next step is to generate a representation of the relevant path constraints. Since \tool prompts the LLM for input generation, the path can be represented in the original programming language (\eg \texttt{C}, \texttt{C++}, or \texttt{Java}), which the LLM can directly reason over. Our approach differs from traditional concolic execution engines, which translate the path into a new representation (\eg an SMT formula) suitable for solving. 

\tool designs a modified {\em dynamic back slicing} algorithm over a concrete execution trace (shown in \autoref{alg:slicing}). The algorithm takes the roadblock $\mathcal{B}$ $\langle \mathit{condStmt}, \mathit{branchCovd}, \mathit{seed} \rangle$ obtained from the previous step. It constructs the AST for the source code (line 3) in advance. It performs a concrete execution using the selected $\mathit{seed}$ to get the execution trace that executes $\mathit{condStmt}$ at least once (line 4). The algorithm then generates a modified dynamic slice along the concrete execution trace as path constraints, suitable for LLM prompting.

It then initializes a set of statements of interest $\mathit{stmts}$ (line 5) and variables of interest $\mathit{vars}$ (line 6). The original $\mathit{condStmt}$ is negated by replacing it with the assertion: $\texttt{assert(!}\mathit{branchCovd}\texttt{)}$. This assertion serves as the {\em target} for the LLM-based concolic execution, which must be held in order to cover new branches.
Specifically, given ($\texttt{if~(}c\texttt{)}~\texttt{then}~...~\texttt{else}~...$), then:
\begin{itemize}[leftmargin=*]
\item[-] $\texttt{assert(!}c\texttt{)}$ is generated if the \texttt{then}-branch is covered;
\item[-] $\texttt{assert(}c\texttt{)}$ is generated if the \texttt{else}-branch is covered.
\end{itemize}
For ($\texttt{switch~(}x\texttt{)~\{~case}~v_1\texttt{:}~...~\texttt{case}~v_n\texttt{:}~...~\texttt{default:}~...~\texttt{\}}$), the assertion is generated as:
{\setlength{\abovedisplayskip}{4pt}%
\setlength{\belowdisplayskip}{4pt}%
\begin{align*}
    \texttt{assert(}x = w_1~\texttt{||}~...~\texttt{||}~x = w_m~\texttt{||}~ \mathit{dflt}\texttt{)}
\end{align*}}
where (1) the subset $\{w_1, ..., w_m\} \subseteq \{v_1, ..., v_n\}$ corresponds to the non-covered \texttt{case}-branches from the \texttt{switch}-statement, and (2) $\mathit{dflt}$ will be the expression $(x \neq v_1~\texttt{\&\&}~...~\texttt{\&\&}~x \neq v_n)$ if the \texttt{default}-branch both exists and is non-covered, or $\mathit{dflt}$ will be $\mathit{false}$ (unused) otherwise.

Once these are built, the modified dynamic slice is constructed. It iterates over the execution trace in reverse order, and collects all statements $\mathit{stmts}$ that $\mathit{condStmt}$ is (transitively) data dependent (lines 7--12). It then accumulates the $\mathit{slice}$ from the data-dependent statements $\mathit{stmts}$ (lines 13--15). This essentially accumulates fragments of the original source file(s) into a coherent slice. For each statement $s \in \mathit{stmts}$, it accumulates $s$ as well as any {\em context} of $s$. Broadly, the context includes all elements of the AST:
\begin{enumerate}[leftmargin=*]
\item control-flow structure enclosing $s$, including \texttt{if}-, \texttt{while}-, \texttt{for}-, and \texttt{switch}- statements;
\item enclosing function definition, including types/parameters;
\item declarations (local and global) for each variable used by $s$;
\item syntax necessary to parse any of the above,
      including braces (\verb+{+/\verb+}+) and semicolons (\verb+;+);
\item formatting (whitespace) and comments.
\end{enumerate}
Finally, the collected set of file ranges in {\em flattened} into a string suitable for prompt generation, preserving the original file formatting and ordering (line 16).


\subsection{LLM-based Input Generation} \label{sec:test_generate}

{
\begin{figure}[t]
  \centering
  \includegraphics[page=1, trim= 0.25in 7.4in 1.3in 0.35in, clip, scale=0.85]{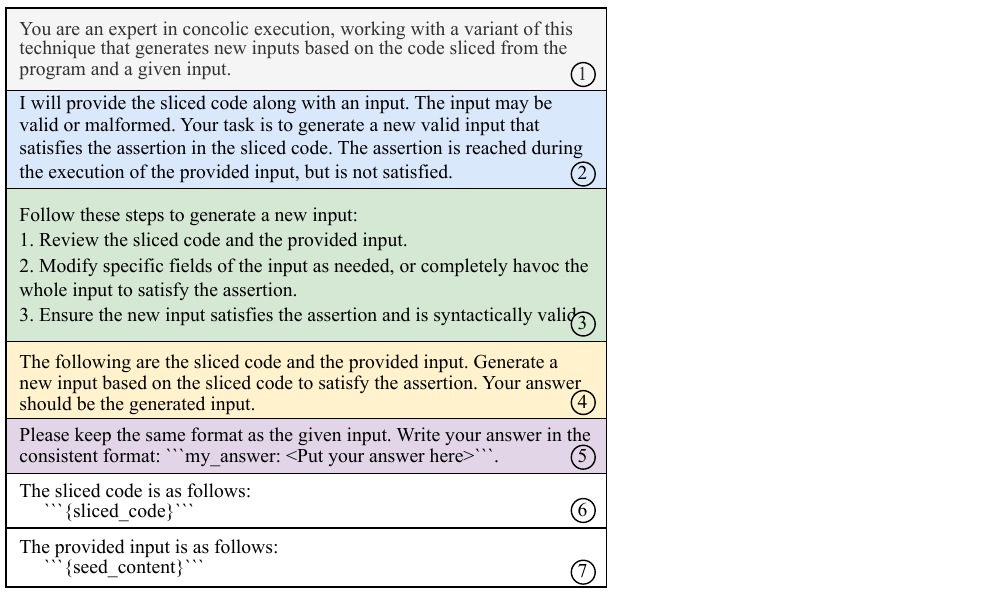}
  \caption{Prompt template for generating new input based on the given input and relevant code slice.}
  \label{fig:prompt}
\end{figure}
}

After collecting the relevant code slice, the LLM is prompted to generate a new syntactically valid input aimed at covering the previously uncovered branch for the roadblock condition. The template of the prompt is shown in \autoref{fig:prompt}, which consists of several parts (\ccc{1}--\ccc{7}). 

The parts \ccc{1}, \ccc{2}, and \ccc{3} together form the system prompt. Part \ccc{1} defines the LLM's persona as an expert in concolic execution, working with a variant of this technique. Based on our observations, the LLMs have sufficient knowledge regarding concolic execution. Since \tool simulates traditional concolic execution, Part \ccc{1} can help the LLM understand the overall task. Part \ccc{2} gives a more detailed description of the task: generating a new syntactically valid input that satisfies the assertion in the sliced code. The whole task of \tool is relatively complex and includes several steps. To maintain the effectiveness of the LLM, Part \ccc{3} includes detailed instructions that break down the task into simpler subtasks, guiding the LLM step by step. While generating a new input, the LLM can slightly modify or havoc the given input, but ensure that it is syntactically valid.  

The parts \ccc{4}--\ccc{7} are included in the user prompt. Part \ccc{4} describes the task and highlights that the answer is a concrete input instead of other formats (e.g., one program that generates inputs). This part also prevents the LLM from printing intermediate results, helping reduce token usage. Part \ccc{5} further specifies the expected output format, which simplifies response extraction and ensures that the generated inputs are immediately usable. Parts \ccc{6} and \ccc{7} include the code slice and the given input, respectively, each enclosed within delimiters for clear separation. The input given is one that reaches the condition but fails to cover the target roadblock branches. Our insight is that slightly modifying a few bytes can often solve the roadblock, which can be simpler than generating a new input from scratch. However, if slight modifications cannot achieve the goal, causing havoc to the input becomes necessary, although this is a secondary option. Furthermore, the input provided serves as an example to illustrate the correct input format (\ie ``one-shot'' prompt) and offers more context to help the LLM reason through the task.

After receiving the input generated by the LLM, \tool adds it to the seed corpus, where it can be automatically retrieved and executed by the greybox fuzzer. The greybox fuzzer is configured as a distributed fuzzer, with a key feature being its ability to analyze seeds from one or more collaborators. Before the generated input is added to the seed corpus, there is no additional validation; instead, this task is delegated to the greybox fuzzer. The fuzzer executes the new input and saves it in the seed corpus if it covers new branches. 
\section{Evaluation} \label{sec:evaluation}
 
To evaluate the efficacy of \tool, we seek to answer the following research questions:

\begin{description}
\item [\textbf{RQ.1}] \textbf{(Branch Coverage and Bug Finding)} Can \tool cover more code branches compared to the baselines? Can \tool find additional bugs over the baselines?\vspace{0.1cm}

\item [\textbf{RQ.2}] \textbf{(Time Usage of Concolic Execution)} How much time does LLM-based concolic execution in \tool take to generate a single input? Is it faster than the baselines?\vspace{0.1cm}

\item [\textbf{RQ.3}] \textbf{(Coverage Improvement Analysis)} How does \tool achieve higher code coverage than the baselines? Does the code coverage improvement stem from solving a larger number of roadblocks? Do the efficiency and the solving capability of LLM-based concolic execution contribute to the code coverage improvement? Does the hybrid architecture of LLM-based concolic execution contribute to the improvement? \vspace{0.1cm}

\item [\textbf{RQ.4}] \textbf{(Unsolved Roadblock Analysis)} How does \tool fail to solve certain roadblocks? How many roadblocks are selected for solving yet remain unsolved in \tool? Does the number of solving attempts affect the results? Are there any roadblocks that are solved by the baselines but remain unsolved by \tool?
\vspace{0.1cm}

\item [\textbf{RQ.5}] \textbf{(Case Study)} Can \tool expose the bug in \texttt{Jenkins} implemented in a different programming language released from the 2024 DARPA and ARPA-H's Artificial Intelligence Cyber Challenge?
\end{description}

\subsection{Experimental Setup} \label{sec:exp_setup}

\subsubsection{{Comparison Baselines}} As baseline tools, we selected three state-of-the-art hybrid fuzzers for comparison: \qsym \cite{qsym}, \intriguer \cite{intriguer}, and \cofuzz \cite{hybrid}. \driller \cite{driller} is the first hybrid fuzzer that combines greybox fuzzing with concolic execution; however, it does not scale to large real-world benchmark subjects, as reported in previous works \cite{qsym, hybrid}. Among existing hybrid fuzzers, only \qsym and \intriguer align with \tool in their goal of improving concolic execution integrated in hybrid fuzzing, although they do so by incorporating fast symbolic emulation and constraint-solving strategies. \cofuzz is the most recent hybrid fuzzer, developed after comprehensively evaluating all previous greybox fuzzers \cite{digfuzz, meuzz, pangolin}, which have also shown lower effectiveness compared to \cofuzz. Each of these hybrid fuzzers integrates its own concolic execution engine with the greybox fuzzer \afl. For a fair comparison, we also integrated \afl into \tool to build a hybrid fuzzer. 

\subsubsection{{Benchmark Subjects}}

\begin{table}[t]
    \caption{Detailed information about benchmark subjects used in our experiments for RQ.1--RQ.4.}
    \label{tab:subjects}
    \centering
    \small
    \setlength\tabcolsep{2.5pt}
    \def\arraystretch{1.15}
    \begin{tabular}{l|rr|l}
        \toprule
        {\bfseries Subject} & {\bfseries \#LOC} & {\bfseries Version} & {\bfseries Description} \\
        \hline
        \hline
        as-new & 96.78k & v2.43 & Assemble code into object files \\
        bc & 11.32k & v1.08.2 & Perform mathematical calculations \\
        calc & 62.03k & v2.16.0.0 & Perform advanced calculations \\
        cflow & 37.39k & v1.7.90 & Generate call graphs for C code \\
        cJSON & 17.23k & v1.7.18 & Parse JSON files \\
        curl & 188.4k & v8.13.0 & Transfer data with URL syntax \\
        cxxfilt & 96.78k & v2.43 & Demangle C++ symbol names \\
        infotocap & 100.48k & v6.4 & Convert terminfo into termcap \\
        jq & 34.11k & a8f27cc & Parse JSON files \\
        libxml2 & 201.95k & v2.13.4 & Parse XML files \\
        Lua & 35.22k & v5.4.7 &  Interpret Lua scripts \\ 
        MuJS & 16.77k & v1.3.5 & Interpret Javascript files \\
        php & 2123.39k & v8.4.6 & Interpret PHP files \\
        SQLite3 & 593.90k & 3.48.0 & Manage SQL database files \\
        \bottomrule
    \end{tabular}
\end{table} 

\autoref{tab:subjects} lists the benchmark subjects used in our evaluation to answer the research questions RQ.1 to RQ.4. Our benchmark comprises 14 widely-used real-world programs, including three parsers (for \texttt{JSON} and \texttt{XML}), two binutils utilities (\verb+as-new+ and \verb+cxxfilt+), three language interpreters (for \texttt{Lua}, \texttt{PHP} and \texttt{JavaScript}), two transformation tools (for \texttt{URL} and \texttt{terminfo}), one GNU utility (\verb+cflow+), one database engine (\verb+SQLite3+), and two mathematical calculators (\verb|bc| and \verb|calc|). These subjects involve both string-based and arithmetic path constraints. 

\begin{table*}[t]
    \caption{Average number of branches covered by our \tool, and the baselines \cofuzz, \intriguer, and \qsym. }
    \label{tab:cov_summary}
    \centering
    \small
    \setlength\tabcolsep{3pt}
    \def\arraystretch{1.15}
    \begin{tabular}{l|r|rrr|rrr|rrr}
        \toprule
        \multirow{2}{*}{ \bfseries  Subject} & \multirow{2}{*}{ \bfseries  \toolBold} &  \multicolumn{3}{c|}{ \bfseries  Comparison with \cofuzzBold} &  \multicolumn{3}{c|}{ \bfseries Comparison with \intriguerBold}  &  \multicolumn{3}{c}{ \bfseries Comparison with \qsymBold}  \\ 
        \cline{3-11} 
         & & { \cofuzzBold}  & { \bfseries Improv} &  { \bfseries SpeedUp} & { \intriguerBold}  & { \bfseries Improv} &  { \bfseries SpeedUp} & { \qsymBold}  & { \bfseries  Improv} &  { \bfseries SpeedUp}   \\
        \hline
        \hline
        as-new & 8145.2  &	6150.0	& 32.44\% & 4.85$\times$ & 5640.2 &	44.41\%	& 6.70$\times$	& 4831.9	& 68.57\%	& 25.71$\times$  \\

        bc & 1203.7 & 984.1 & 22.31\% & 5.49$\times$ & 1015.1 & 18.58\%	& 5.40$\times$ & 982.3 & 22.54\%	& 5.49$\times$\\

        calc & 7194.5 & 4724.3 & 52.29\% & 18.49$\times$ & 5328.1 & 35.03\% &	5.74$\times$ & 4795.1 & 50.04\% &	17.36$\times$\\
        
        cflow & 1238.2	&	1167.2	& 6.08\% & 11.16$\times$ & 1194.1&	3.69\%&	6.43$\times$	&1152.9&	7.40\%&	14.55$\times$\\
        
        cJSON & 379.9&	375.5	& 1.17\% & 22.50$\times$ & 347.4&	9.36\%	&57.60$\times$	&343.9&	10.47\%	&80.00$\times$ \\

        curl & 3789.0 & 3285.0 &	15.34\%	& 6.75$\times$	&	3379.0	& 12.13\% &	1.35$\times$	&	3281.0	& 15.48\%	& 6.76$\times$ \\
        
        cxxfilt & 2448.4	& 2199.6	& 11.31\% &2.78$\times$ &2039.1&	20.07\%	&3.80$\times$	&1951.0&	25.49\%&	4.39$\times$ \\

        infotocap & 1729.0 & 1545.4	& 11.88\%	& 2.84$\times$ & 	1474.9	& 17.23\%	& 3.72$\times$	&	933.2	& 85.28\%	& 240.00$\times$ \\

        jq & 2076.0	& 2026.2	& 2.46\%	& 30.64$\times$ &	2039.7	& 1.78\%	& 27.17$\times$	&	2018.6 &	2.84\%	& 34.29$\times$\\
        
        libxml2 & 6811.6& 3282.4 & 	107.52\% &32.00$\times$ & 3163.8&	115.30\%	&49.66$\times$&	3092.3&	120.28\%	&51.43$\times$ \\
        
        Lua & 3529.3& 2923.8	& 20.71\%	& 5.45$\times$ & 2438.0	& 44.76\% & 12.20$\times$ & 1951.5	& 80.85\% & 24.83$\times$ \\
        
        MuJS & 3666.0&	2454.0	& 49.39\% & 7.70$\times$& 2078.0&	76.42\%	&14.85$\times$	&2096.0	&74.90\%	&14.69$\times$  \\

        php & 17268.7 &	8926.5	& 93.45\%	& 6.26$\times$	&	5554.2	& 210.91\%	&53.33$\times$	&	5683.5	& 203.84\%	&45.00$\times$\\
        
        SQLite3 & 19746.0 & 17369.5 &	13.68\% & 1.75$\times$ & 17300.3&	14.14\% & 1.76$\times$ & 11990.0 & 	64.69\% & 96.00$\times$ \\
        \hline
        \hline
        \textbf{Average} & - & -	& 31.43\% & 11.33$\times$ & - & 44.56\%	& 17.84$\times$	&  - & 59.48\% & 47.18$\times$ \\
        \bottomrule
    \end{tabular}
\end{table*} 

For each subject, we report the details on the code lines, the version used in our evaluation, and their primary functionality. These subjects are widely utilized in practice and represent diverse functionalities and codebase sizes. Our benchmark includes all text-based input subjects that are incorporated into popular fuzzing benchmark suites: \magma \cite{Magma}, \unifuzz \cite{unifuzz}, and \fuzzbench \cite{fuzzbench}, as well as all those used in existing hybrid fuzzing works \cite{qsym, intriguer, hybrid}. In addition to these string-heavy programs, we include two widely-used numerical-heavy programs (i.e., \verb|bc| and \verb|calc|) to evaluate the generality of \tool. 

\subsubsection{{Configuration Parameters}} For the subject programs, we adopted the same configuration parameters as those used in previous work \cite{fuzzbench, Magma, unifuzz, hybrid}. We used the \verb+gpt-4o+ model as the large language model, with the temperature set to 0.5 for precise and factual responses. Additionally, to determine if the greybox fuzzer is stuck, we set a time threshold of 90 seconds; if the fuzzer fails to cover new code paths within this period, the concolic execution engine is triggered. This time frame aligns with the default settings used by the baseline tools. Following suggestions from the fuzzing community \cite{bohme2020fuzzing, evaluating}, each experiment was run for 24 hours, and we report the average over 10 runs to mitigate the impact of randomness. All experiments were conducted on an Intel{\textregistered} Xeon{\textregistered} Platinum 8468V CPU with 192 logical cores clocked at 2.70GHz, 512GB of memory, and Ubuntu 22.04.3 LTS. 

\subsection{Branch Coverage and Bug Finding (RQ.1)} \label{sec:req1}

The primary goal of hybrid fuzzers is to maximize code coverage, making it a key metric for evaluating the performance of hybrid fuzzers. In our evaluation, we measured the branch coverage achieved by \tool and the state-of-the-art hybrid fuzzers using \verb|gcovr|. To ensure a fair comparison, we used the same seeds for each fuzzer. Each fuzzer was run on each subject for 24 hours and repeated 10 times; we report the average data (i.e., mean values) as the final results. Since higher code coverage typically increases the likelihood of exposing bugs \cite{evaluating}, we also report the number of bugs discovered during these runs. 

\subsubsection{Results on Branch Coverage}

\autoref{tab:cov_summary} shows the final code coverage achieved by \tool compared to three baseline fuzzers. We report the average number (i.e., mean values) of branches covered by each fuzzer in 24 hours across 10 runs. To quantify the improvement of \tool over the baseline tools, we report the percentage increase in terms of branch coverage in 24 hours (\emph{Improv}), and how much faster \tool can achieve the same branch coverage as the baseline tools (\emph{SpeedUp}). 

Overall, \tool achieved more branch coverage than all state-of-the-art hybrid fuzzers on all subjects, with improvements ranging from 1.17\% to 210.91\%. Compared to \cofuzz, \intriguer, and \qsym, \tool covered 31.43\%, 44.56\%, and 59.48\% more code branches on average, respectively. In addition, \tool covered the same number of branches 11.33$\times$ faster than \cofuzz, 17.84$\times$ faster than \intriguer, and 47.18$\times$ faster than \qsym. 

To account for randomness, we measured the Vargha-Delaney \cite{vargha-delaney} effect size ${\hat{A}_{12}}$, which indicates the probability that a random campaign of \tool outperforms a random campaign of comparison tools. In all subjects, ${\hat{A}_{12}} = 1$ (that is, ${\hat{A}_{12}} \geq 0.70$), indicating that \tool consistently achieved higher branch coverage in all runs than each baseline. This shows a significant improvement of \tool over baselines in code coverage.

In addition, we measured both the monetary costs and the token usage required to run these experiments. It is worth noting that the monetary cost is relatively low: in our experiments, running a subject for one hour only cost 0.40 USD on average. The required token count was also modest. The average token count per prompt was 1,570, with the maximum observed being 3,020 tokens in \verb|php| containing millions of code lines. We did not encounter cases where the generated code slices exceeded the LLM's context window limit. These results further demonstrated the practical utility of \tool. 

\subsubsection{Results on Bug Finding} Although these subjects have been extensively tested by existing fuzzers, \tool found seven unique and previously unknown bugs in their latest version: two in \verb+cxxfilt+, two in \verb+php+, and three in \verb|calc|. In contrast, the baseline fuzzers only found two (one in \verb+php+ and one in \verb|calc|) of these bugs in total, and also did not find any additional bugs. We reported these bugs to the respective developers, and all have been confirmed. These bugs include null pointer dereference, undefined behavior, and buffer overflows. 

Upon closer investigation, we found that the missed bugs were hidden behind complex path constraints. For example, exposing the bugs in \verb|cxxfilt| requires solving deep path constraints---involving 13 and 16 function calls, respectively. The baselines faced a similar difficulty in exposing the bugs in \verb|calc|. Concolic execution engines in the baselines often drop some path constraints for better solving efficiency, but also compromise the performance. We observed that the solving capability of the baselines significantly declined with the increase in the path constraint depth. This observation is consistent with the finding of a recent study \cite{wu2024tumbling}. In addition, exposing the \verb|php| bug missed by the baselines requires generating a valid function in the \texttt{PHP} input: \texttt{function name()\{\}}, where each byte corresponds to one branch condition. Generating this segment requires multiple rounds of constraint computing and solving. Traditional concolic execution just managed to generate a partial segment (e.g., \texttt{fun}), thereby failing to trigger the bug. 

\result{\tool achieves significantly higher code coverage than \cofuzz, \intriguer, and \qsym, covering 31.43\%, 44.56\%, and 59.48\% more code on average, respectively. Moreover, \tool achieves the same number of branches 11.33 times, 17.84 times, and 47.18 times faster than these baselines. In extensively tested projects, \tool exposed seven unique and previously unknown bugs, while baselines only found two of them without finding any additional bugs.}

\subsection{Time Usage of Concolic Execution (RQ.2)} \label{sec:req3}

To evaluate the efficiency of LLM-based concolic execution, we measured the time required for each concolic execution run to generate an input. Specifically, for the baselines \intriguer, \qsym, and \cofuzz, we measured the time used for constraint computing alongside concrete execution and constraint solving via SMT solvers to generate an input. Similarly, for \tool, we measured the time used for dynamic code slicing along the concrete execution, and constraint solving using an LLM to generate an input---from the prompt construction to obtaining a concrete input from the LLM.  

{
\begin{figure}[t]
  \centering
  \begin{minipage}[t]{\linewidth}
    \centering
    \includegraphics[page=4, trim=0.2in 5.45in 9.0in 0.1in, clip, scale=0.82]{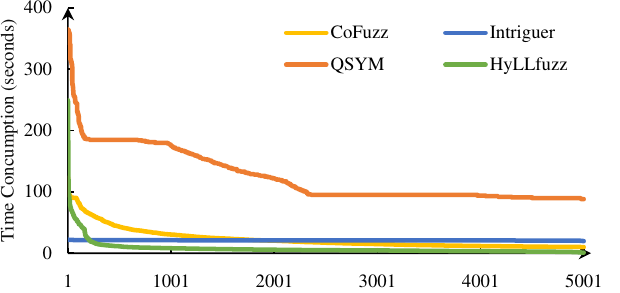}
    \caption*{(a)}
  \end{minipage}
  \\[7pt]
  \begin{minipage}[t]{\linewidth}
    \centering
    \small
    \setlength\tabcolsep{1pt}
    \def\arraystretch{1.2}
    \begin{tabular}{c|cr|cr|cr}
      \toprule
      { \toolBold} & \multicolumn{2}{c|}{ \textbf{\textit{vs}} \cofuzzBold} & \multicolumn{2}{c|}{ \textbf{\textit{vs}} \intriguerBold} & \multicolumn{2}{c}{ \textbf{\textit{vs}} \qsymBold} \\
      \hline
      \hline
     { \textbf{Med}} & { \textbf{Med}} & { \textbf{Speedup}} & { \textbf{Med}} & { \textbf{Speedup}} & { \textbf{Med}} & { \textbf{Speedup}} \\
      \hline
      4.97 & 17.39 & 3.50$\times$ &  20.95 & 4.22$\times$ & 95.00 & 19.11$\times$ \\
      \hline
      \hline
      { \textbf{Avg}} & { \textbf{Avg}} & { \textbf{Speedup}} & { \textbf{Avg}} & { \textbf{Speedup}} & { \textbf{Avg}} & { \textbf{Speedup}} \\
      \hline
      7.66 & 23.72 & 3.10$\times$ & 20.96 &  2.74$\times$ & 125.26 & 16.35$\times$\\
      \bottomrule 
    \end{tabular}
    \caption*{(b)}
  \end{minipage}
  \caption{Wall-clock time used by the baselines \cofuzz, \intriguer, \qsym, and our \tool for each run of concolic execution (seconds).}
  \label{fig:time}
\end{figure}
}

We randomly sampled 5000 concolic execution runs of \tool and three baselines \cofuzz, \intriguer, and \qsym across all subjects, respectively. \autoref{fig:time} (a) shows the time usage of each fuzzer within these sampled runs, while \autoref{fig:time} (b) shows the average time and median time usage. As we can see, compared to all these baselines, \tool substantially reduces time usage, although these baselines adopt optimization strategies. 

In this experiment, we reported and compared the wall-clock time used, as it is a standard metric for evaluating fuzzers and is also particularly important for end-users seeking to mitigate security risks. We acknowledge that, due to the use of an LLM, \tool incurs additional GPU resource consumption compared to the baselines; however, we believe this cost is justified by the substantial gains in efficacy. 

\result{Each concolic execution run in \tool takes 4.97 seconds in the median, which is 3.50 times, 4.22 times, and 19.11 times faster than that of \cofuzz,  \intriguer, and \qsym, respectively.}

\subsection{Coverage Improvement Analysis (RQ.3)} \label{sec:req2}


We further conducted an in-depth analysis to understand how \tool achieved a significant improvement in code coverage compared to the baselines. In this analysis, we compared \tool with hybrid fuzzers \cofuzz, \intriguer, and \qsym to evaluate the performance of different concolic execution engines. In addition, we included an additional comparison with the grammar-aware fuzzers \nautilus \cite{nautilus} and \superion \cite{superion} to evaluate whether the LLM-based constraint reasoning and solving contribute to the code coverage improvement. 

\begin{table}[t]
    \caption{Average number of roadblocks solved by our \tool and baselines \cofuzz, \intriguer and \qsym.}
    \label{tab:roadblocks}
    \centering
    \small
    \setlength\tabcolsep{5pt}
    \def\arraystretch{1.15}
    \begin{tabular}{l|r|rrr}
        \toprule
        {\bfseries Subject} & {\toolBold} & {\cofuzzBold} & {\intriguerBold} & {\qsymBold} \\
        \hline
        \hline
        as-new & 585.0 & 298.7 & 250.4 & 155.1 \\
        bc & 200.0 & 74.5 & 135.0 & 70.0\\
        calc & 880.1 & 177.0 & 361.0 & 172.0 \\
        cflow & 100.0 & 55.4 & 59.3 & 26.9 \\
        cJSON & 84.5 & 72.9 & 18.4 & 0.0 \\
        curl & 76.3 & 89.0 & 36.6 & 56.7 \\
        cxxfilt & 193.8 & 139.7 & 106.3 & 104.8\\
        infotocap & 309.0 & 265.0 & 248.1 & 165.0 \\
        jq & 44.0 & 1.3 & 4.4 & 0.1 \\
        libxml2 & 842.0 & 224.0 & 158.1 & 269.0 \\
        Lua & 624.3 & 301.1 & 290.1 & 112.0 \\ 
        MuJS & 758.6 & 454.8 & 315.1 & 378.2\\
        php & 837.1 & 698.3 & 423.0 & 511.0 \\
        SQLite3 & 224.6 & 121.6 & 136.1 & 0.0  \\
        \hline
        \hline
        {\bfseries Average} & 411.4 & 212.4 & 181.6 & 144.3\\
        \bottomrule
    \end{tabular}
\end{table} 

\subsubsection{{Number of Solved Roadblocks}} We started by analyzing whether the increase in branch coverage correlates with the number of roadblocks solved. In \autoref{tab:roadblocks}, we report the average number of roadblocks solved per run by \tool and the baselines based on 10 runs of 24 hours. On average, \tool solved 411.4 roadblocks, while the baselines solved 212.4, 181.6, and 144.3 roadblocks, respectively. In certain subjects (e.g., \verb+jq+, \verb+curl+, and \verb|cJSON|), \tool solved relatively few roadblocks, while \tool solved more than 800 roadblocks in subjects \verb|calc|, \verb|libxml2| and \verb|php|. In all subjects, \tool consistently outperformed all baseline fuzzers in terms of the number of roadblocks solved within the same time frame. When comparing the performance of \tool in different subjects, we observe that a larger number of solved roadblocks almost corresponds to a greater increase in code coverage. These results demonstrate a positive correlation between the number of roadblocks solved and the improvement in code coverage.

\subsubsection{{Contribution of Fast Concolic Execution}} We further analyzed whether the increase in branch coverage is due to the fast concolic execution engine. To this end, we measured the number of inputs generated by the concolic execution engines of \tool and the baselines within 24 hours. On average, \tool generated 2828 inputs, while \cofuzz, \intriguer, and \qsym generated 2494, 2154, and 1475 inputs, respectively. \tool significantly outperformed \qsym, generating more than twice as many inputs in the same time frame. However, compared to \cofuzz and \intriguer, \tool only achieved a slightly higher input number. This result indicates that fast concolic execution contributes to improved code coverage. However, the performance gain cannot be fully attributed to the fast LLM-based engine of \tool, as \cofuzz and \intriguer generated a comparable number of inputs. 

\subsubsection{{Contribution of LLM-based Solving}} We next analyzed whether the solving capability of \tool contributed to increasing code coverage. We examined this from two aspects. First, we used the concolic execution engines of the baselines to solve the roadblocks selected by \tool, and then compared the number of roadblocks solved by different techniques under the same time budget (i.e., 24 hours). This analysis is intended to demonstrate that the performance gains of \tool stem from LLM-based solving rather than from the differences in roadblock selection. Second, we conducted a manual analysis of the branches covered by \tool but missed by the baselines to elucidate the role of LLM-based solving in this process. To this end, we selected two string-based subjects \verb|php| and \verb|jq|, where \tool achieved a higher increase in code coverage in \verb|php| but only a smaller improvement in \verb|jq|, as well as two arithmetic-based subjects \verb|bc| and \verb|calc|. 

\begin{table}[t]
    \caption{Number of roadblocks solved by baselines using \tool-selected roadblocks (\emph{baseline-modified}) and baselines with their original designs (\emph{baseline-original}), and our \tool.}
    \label{tab:ablation}
    \centering
    \small
    \setlength\tabcolsep{9pt}
    \def\arraystretch{1.15}
    \begin{tabular}{l|rrrr}
        \toprule
        {\bfseries Tool}
        & \multicolumn{1}{c}{\bfseries php}
        & \multicolumn{1}{c}{\bfseries jq}
        & \multicolumn{1}{c}{\bfseries bc}
        & \multicolumn{1}{c}{\bfseries calc} \\
        \hline
        \hline
        baseline-modified & 205.3 & 9.0 & 117.0 & 369.0 \\
        baseline-original & 698.3 & 4.4 & 135.0 & 361.0 \\
        \hline
        \tool & 827.1 & 44.0 & 200.0 & 880.1 \\
        \bottomrule
    \end{tabular}
\end{table} 

In \autoref{tab:ablation}, we report the results of using the concolic execution engines of the baselines to solve the roadblocks selected by \tool. The best-performing baselines could only solve 205.3, 9.0, 117.0, and 369.0 roadblocks in \verb|php|, \verb|jq|, \verb|bc|, and \verb|calc|, respectively. These results are similar to those obtained using their original designs, in which the best baseline solved 698.3, 4.4, 135.0, and 361.0 roadblocks, respectively. In contrast, our \tool using LLM-based solving solved substantially more roadblocks, whose numbers are 837.1, 44.0, 200.0, and 880.1 in the same subjects, respectively. These results indicate that the improvement of \tool in code coverage stems from its LLM-based solving ability rather than from the use of a different roadlock-selection strategy. 

{
\begin{figure}[t]
\begin{lstlisting}[language=MyC, frame=single, numbers=left, breaklines=true,
  numbersep=5pt,           
  xleftmargin=15pt,        
  framexleftmargin=12pt,
  numberstyle=\color{black}\ttfamily,
  mathescape=true]
yy523:
  yych = *++YYCURSOR;
  if (yych == 'R') goto yy597;
  if (yych == 'r') goto yy597;
  goto yy140;

yy597:
  ++YYCURSOR;
  if ((yych = *YYCURSOR) <= '^') {...}
  else {
    if (yych <= 'd') { ... } 
    else {
      if (yych <= 'z') goto yy655; ...
    }
  }
\end{lstlisting}
\caption{Code fragment from \texttt{php} showing input interpretation.}
\label{fig:exp_code}
\end{figure}
}

We further manually analyzed the branches covered by \tool but missed by the baselines. \autoref{fig:exp_code} illustrates how \verb|php| interprets keywords containing the byte `\texttt{R}'. In our evaluation, the baselines failed to generate the simple keyword \texttt{try}, which is necessary to explore certain program behaviors. After interpreting the input ``\texttt{tr}'', \tool inferred the code semantics and correctly generated the missing letter ``\texttt{y}'' to solve the condition on line 13, generating a syntactically valid input and achieving global optima. In contrast, the baselines focus only on local optima and prioritize the value ``\texttt{z}'', generating ``\texttt{trz}''. This example demonstrates how \tool generates a single keyword effectively. The format \verb|php| includes more complex keywords like ``\texttt{return}'', ``\texttt{break}'', and ``\texttt{function}''. Without knowledge of the program semantics, the baselines struggled to generate them. Conversely, the \texttt{jq} inputs are \texttt{JSON} files with relatively simple structures; for example, ``\{\}'' represents key-value pairs and ``[]'' represents arrays. This makes \tool just perform slightly better than the baselines in \texttt{jq}. For the arithmetic-based subjects \verb|bc| and \verb|calc|, we observed that the concolic execution engines of the baselines failed to solve the deep path constraints. The success rate of constraint solving significantly decreased as the depth of path constraints increased, which is also reported in the recent study \cite{wu2024tumbling}.

\subsubsection{{Contribution of LLM-based Deeper Reasoning}} From the previous analysis, we observed that one advantage of LLM-based solving is its awareness of input structures compared to traditional SMT-based solvers, which contributes to the coverage improvement to some extent. This raises the following questions: What is the contribution of the \tool's deeper reasoning over the path constraints? Can the improvement be achieved solely through the ability to generate syntactically valid inputs?  

To answer these questions, we included two state-of-the-art grammar-aware fuzzers \nautilus \cite{nautilus} and \superion \cite{superion} for comparison. Grammar-aware fuzzers are specifically designed to handle structured input formats. Such a comparison can highlight the contribution of the deeper reasoning over path constraints during input generation. We compared the branch coverage achieved by \tool with that achieved by grammar-aware fuzzers. \langfuzz \cite{langfuzz} is not included, as its source code is not publicly available. Both \nautilus and \superion rely on input specifications to generate new inputs, and these specifications are manually written. By default, \nautilus only supports the \verb|javascript|, \verb|php|, \verb|lua|, and \verb|xml| inputs, while \superion supports the \verb|javascript| and \verb|xml| inputs. We also attempted to use LLMs to automate specification generation; however, the current LLMs fail to achieve this. As a result, the grammar-aware fuzzers could only fuzz the subjects \verb|libxml2|, \verb|Lua|, \verb|MuJS|, and \verb|php|. We ran each fuzzer on each subject for 24 hours and repeated 10 times.

The average number of branches covered by each fuzzer is shown in \autoref{tab:grammar_fuzzer}. As we can see, \tool covers substantially more code branches than both baselines. Specifically, compared to \nautilus and \superion, \tool covered 66.07\% and 127.04\% more branches, respectively. These results indicate that to achieve this code coverage improvement, only handling the input structures is not sufficient, and the deeper reasoning over path constraints plays a significant role in this improvement.

\begin{table}[t]
    \caption{Average number of branches covered our \tool, and grammar-aware fuzzers \nautilus and \superion. }
    \label{tab:grammar_fuzzer}
    \centering
    \small
    \setlength\tabcolsep{2pt}
    \def\arraystretch{1.15}
    \begin{tabular}{l|r|rr|rr}
        \toprule
        \multirow{2}{*}{ \bfseries  Subject} & {\footnotesize \toolBold} &  \multicolumn{2}{c|}{ \nautilusBold} &  \multicolumn{2}{c}{\superionBold}  \\ 
        \cline{2-6} 
         & {\bfseries \#Branches} & {\bfseries \#Branches}  & { \bfseries Improv} & {\bfseries \#Branches}  & { \bfseries Improv} \\
        \hline
        \hline
        
        libxml2 & 6811.6& 3126.0	& 117.90\%	& 2503.1	& 172.13\% \\
        
        Lua & 3529.3& 2358.7 &	49.63\% & 	-	& - \\
        
        MuJS & 3666.0&	2234.7	& 64.05\%	& 2014.9	& 81.94\%  \\

        php & 17268.7 &	13011.5	& 32.72\%	& - & -	 \\
        \hline
        \hline
        \textbf{Average} & - & -	& 66.07\% & - & 127.04\% \\
        \bottomrule
    \end{tabular}
\end{table} 

\subsubsection{{Contribution of Hybrid Architecture of LLM-based Concolic Execution}} Concolic execution adopts a hybrid architecture that combines concrete execution with symbolic execution. Our LLM-based concolic execution approach follows this same paradigm. However, given that LLMs have the potential of reasoning over concrete execution traces and generating inputs, it is natural to question whether this hybrid architecture remains necessary. In other words, could an LLM combined solely with concrete execution already achieve comparable results? 

\begin{table}[t]
    \caption{Average number of branches covered by both \tool and its ablation tool \tool without only concrete execution.}
    \label{tab:hybrid-coverage}
    \centering
    \small
    \setlength\tabcolsep{9pt}
    \def\arraystretch{1.15}
    \begin{tabular}{l|rrr}
        \toprule
        {\bfseries Subject}
        & \tool
        & \ablation
        & Improv \\
        \hline
        \hline
        php & 17268.7 & 12656.2 & 36.44\% \\
        jq  & 2076.0  & 2023.1  & 2.61\% \\
        bc  & 1203.7  & 1031.4  & 16.71\% \\
        calc& 7194.5  & 5087.6  & 41.41\%  \\ 
        \hline
        \hline 
        {\bfseries Average} & - & - & 24.29\% \\
        \bottomrule
    \end{tabular}
\end{table} 

To investigate this question, we developed an ablation baseline \ablation, which uses the LLM directly to reason over concrete execution and then generate new inputs, while keeping all other strategies (e.g., roadblock selection) the same. We evaluated \ablation on the same set of subjects \texttt{php}, \texttt{jq}, \texttt{bc}, and \texttt{calc} as that used in our previous analysis. Each subject was run for 24 hours and repeated 10 times. The average code coverage achieved by both \tool and its ablation \ablation is shown in \autoref{tab:hybrid-coverage}. From this result, we could see that the hybrid architecture is necessary, as it significantly improves the code coverage. In addition, \tool also reduces the LLM token usage by approximately 3.5$\times$ compared to \ablation. Therefore, this hybrid architecture not only improves coverage but also lowers monetary costs.   

\result{The code improvement of \tool stems from its ability to solve a larger number of roadblocks compared to the baselines. The efficiency of LLM-based concolic execution, the LLM-based reasoning and constraint-solving capabilities, and the hybrid architecture of LLM-based concolic execution all contribute to this improvement.}

\subsection{Unsolved Roadblock Analysis (RQ.4)} \label{sec:req4}

In this analysis, we evaluated how \tool failed to solve certain roadblocks, which aims to provide insights about how to effectively use LLM-based concolic execution. Specifically, we analyzed how many roadblocks are selected for solving but remain unsolved by \tool. We further investigated whether the number of solving attempts affects the results. In addition, we evaluated whether there are roadblocks that are solved by the baselines but not solved by \tool. These analyses are conducted on two string-based subjects \verb|jq| and \verb|php|, and two algorithm-based subjects \verb|bc| and \verb|calc|. 

\subsubsection{Number of Roadlocks Selected for Solving yet Unsolved by \tool} During the fuzzing campaign, we tracked how many roadblocks were selected by \tool for solving, and how many roadblocks were unfortunately unsolved by \tool at the end. The corresponding statistics and unsolved rates are reported in \autoref{tab:unsolved_roadblocks}. Overall, \tool successfully solved more than 60\% of the selected roadblocks, while the unsolved rates range from 16.04\% to 39.64\% across the evaluated subjects. 

We then manually analyzed the reasons why \tool failed to solve these roadblocks. It is surprising to find that this is not due to the limitations of LLM-based concolic execution. Instead, the majority of the roadblocks (up to 80\%) left unsolved by \tool are in fact unsatisfiable, as the corresponding branches are controlled by their configurations and program environments rather than by the fuzzed inputs. For example, a branch condition in \verb|calc| checks whether logging information should be printed. Since printing is disabled in the configuration, the flipped branch cannot be satisfied. Similarly, for the branch conditions that check whether the files are successfully opened or whether memory allocation succeeds, their flipped branches are also unsatisfiable, as they depend on the execution environments. In the current design of \tool, roadblocks are selected uniformly for solving without performing any satisfiability analysis, and branches that are inherently unsatisfiable are not deprioritized. A more effective design would consider the feasibility of a roadblock before selection; we leave this direction for future work.      

\begin{table}[t]
    \caption{Number of roadblocks selected for solving (\emph{\#Selected}), number of roadblocks finally unsolved by \tool (\emph{\#Unsolved}), and the unsolved rate (\emph{UnsolvedRate}).}
    \label{tab:unsolved_roadblocks}
    \centering
    \small
    \setlength\tabcolsep{9pt}
    \def\arraystretch{1.15}
    \begin{tabular}{l|rrrr}
        \toprule
        {\bfseries Tool}
        & \multicolumn{1}{c}{\bfseries php}
        & \multicolumn{1}{c}{\bfseries jq}
        & \multicolumn{1}{c}{\bfseries bc}
        & \multicolumn{1}{c}{\bfseries calc} \\
        \hline
        \hline
        \#Selected & 1131.2 & 72.9 & 238.2 & 1091.1  \\
        \#Unsolved & 294.1 & 28.9 & 38.2 & 211.0 \\
        \hline
        UnsolvedRate   & 26.26\% & 39.64\% & 16.04\% & 19.34\% \\
        \bottomrule
    \end{tabular}
\end{table} 

\subsubsection{Number of Solving Attempts Required to Solve a Roadlock} Given the nondeterministic nature of the LLM responses, we further analyzed whether the number of solving attempts affects the success rate of solving. To this end, we counted the number of solved roadblocks under different numbers of solving attempts, and the results are illustrated in \autoref{fig:solving_times}. We observe that more than 50\% of the roadblocks are solved with a single solving attempt, and that using fewer than five solving attempts is sufficient to solve 94\% of all solvable roadblocks. These results indicate that the LLM responses are relatively stable, and that increasing the number of solving attempts up to five can improve the success rate; however, additional attempts beyond five yield little further improvement. 

{
\begin{figure}[t]
  \centering
  \includegraphics[page=1, trim= 0.1in 4.2in 8.5in 0in, clip, scale=0.73]{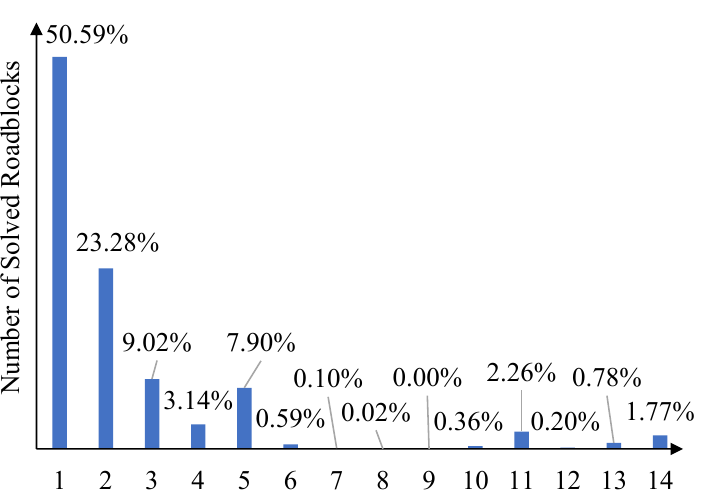}
  \caption{Number of solved roadblocks (in percentage) under different numbers of solving attempts.}
  \label{fig:solving_times}
\end{figure}
}

\subsubsection{Number of Roadlocks Solved by Baselines but Unsolved by \tool} We conducted an additional analysis to evaluate whether there are roadblocks that are solved by the baselines but remain unsolved by \tool. To this end, we counted the code branches uniquely covered by \tool and by the baselines. The results are shown in \autoref{fig:overlap}. As expected, \tool uniquely covered substantially more branches, reflecting its significant coverage improvement. In contrast, the three baselines \cofuzz, \intriguer, and \qsym also uniquely covered a small number of branches in the subjects: 8 branches in \verb|bc|, 144 branches in \verb|calc|, 0 branches in \verb|jq|, and 556 branches in \verb|php|. 

We then manually analyzed why these branches were not covered by \tool. While we initially expected to reveal limitations of LLM-based concolic execution used in \tool, we found that this issue does not arise from its inability to solve the relevant path constraints. Instead, most of the branches uncovered by \tool occur in \verb|switch-case| conditions, where multiple branches coexist. In such cases, \tool represents the solving target in the \emph{or} form after flipping the conditions, and then randomly selects one path conditions to satisfy. As a result, some branches are missed, because \tool chooses one constraint to satisfy while the baselines solve a different path constraint. This explains why \tool ultimately fails to cover some branches.

\result{\tool failed to solve approximately 16.04\% to 39.64\% of the selected roadblocks. While increasing the number of solving attempts can slightly improve the success rate, this effect is limited to five attempts. Additionally, a small number of branches were solved by the baselines but missed by \tool. Our manual analysis revealed that these unsolved branches are not due to limitations of LLM-based concolic execution in \tool.}

{
\begin{figure}[t]
  \centering
  \includegraphics[page=1, trim= 0.1in 0in 4in 0.4in, clip, scale=0.37]{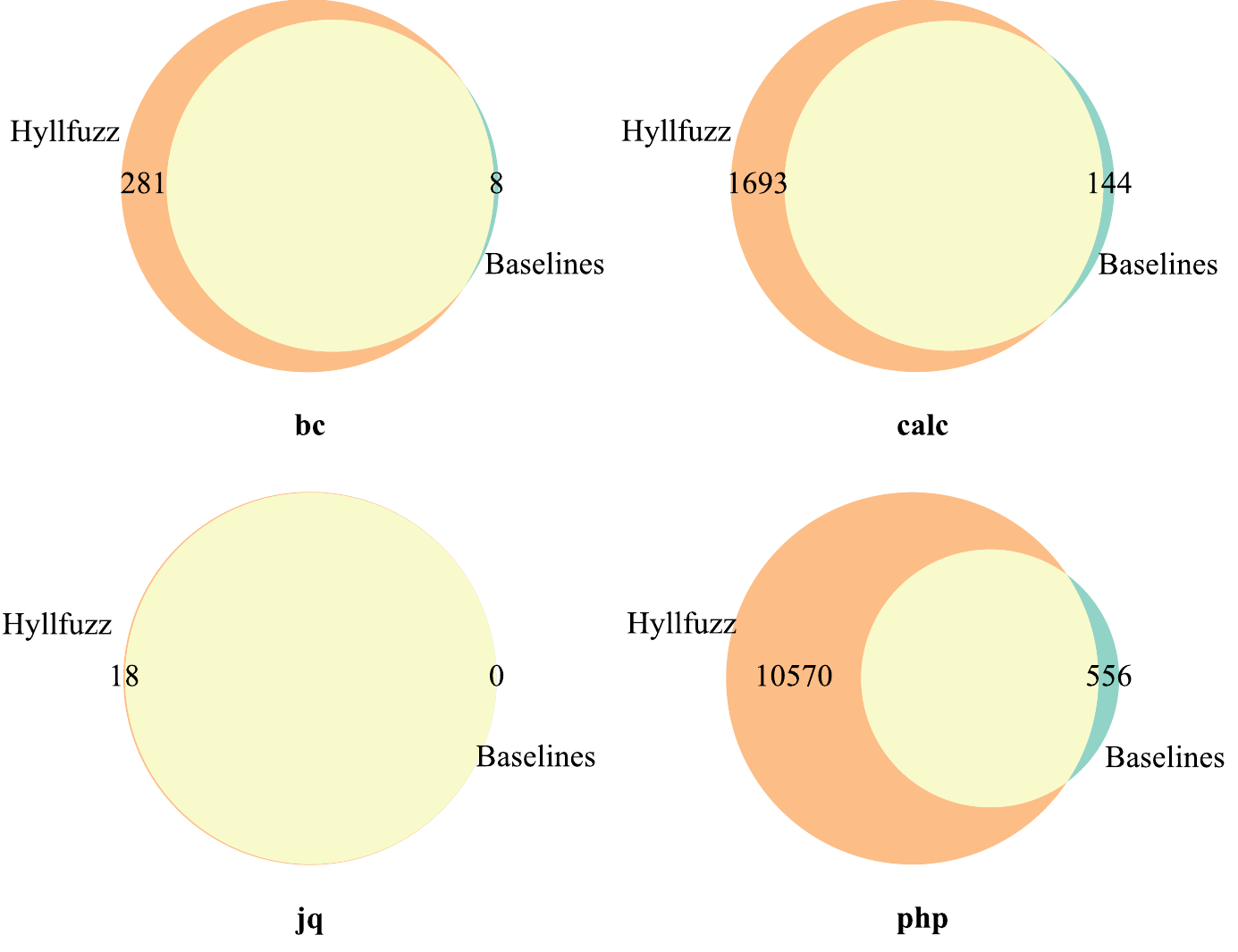}
  \caption{Number of code branches uniquely covered by \tool (in orange) and the baselines (in blue).}
  \label{fig:overlap}
\end{figure}
}

\subsection{Case Study (RQ.5)} \label{sec:req5}

\begin{figure}[t]
  \centering
  \includegraphics[page=1, trim= 0.27in 0.32in 0.27in 0.37in, clip, scale=0.85]{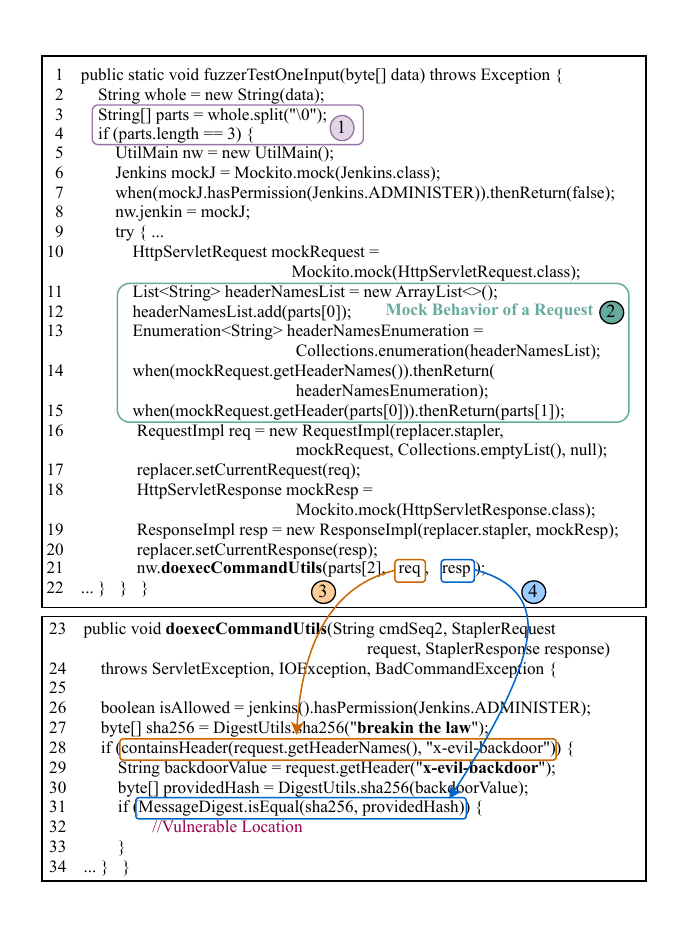}
  \caption{Case study of the bug from the Jenkins subject in Java released in the 2024 DARPA and ARPA-H's AIxCC.}
  \label{fig:case_study}
\end{figure}

In the previous research questions, we evaluated the performance of \tool on \texttt{C}/\texttt{C++} programs in combination with \afl. \tool has also been adapted for testing programs written in other programming languages and combined with other greybox fuzzers. We conducted a case study on the Jenkins subject \cite{jenkins} written in \texttt{Java}, by integrating the greybox fuzzer \jazzer \cite{jazzer} (designed to fuzz Java programs) into \tool. For this study, we selected the bug manually created by the 2024 DARPA and ARPA-H's Artificial Intelligence Cyber Challenge (AIxCC) \cite{aixcc}, and released after the \texttt{gpt-4o} snapshot. The details are shown in \autoref{fig:case_study}. 

The bug occurs on line 32 in \autoref{fig:case_study}. To reach this location, the input must satisfy several conditions. In condition \ccc{1}, the input data is split by the symbol ``\texttt{\textbackslash 0}'', and the resulting parts must have a length of 3. Once this condition is satisfied, the input passes the first constraint and triggers the execution of code \ccc{2}. Although no new conditions are introduced in \ccc{2}, it mocks the behavior of a request, and any concolic execution engines require proper modeling of these behaviors. Following this, the function \texttt{doexecCommandUtils} is invoked, with the three input parts passed as its arguments: \texttt{parts[2]}, \texttt{parts[0]} for \texttt{req}, and \texttt{parts[1]} for \texttt{resp}. Upon entering this function, to satisfy the condition \ccc{3}, the request value (i.e., \texttt{parts[0]}) must contain the string ``\texttt{x-evil-backdoor}''. The return value is then assigned to the variable \texttt{backdoorValue} (line 29), which is derived from \texttt{parts[1]} (line 15). This value is hashed using the \texttt{sha256} function (line 30), and the resulting hash must match the hash of ``\texttt{breakin the law}'' to satisfy the final condition \ccc{4}.

To expose the bug, the greybox fuzzer generated inputs that reached the condition at line 28; however, it failed to generate correct inputs to cover the branch. At this point, \tool was invoked to identify this roadblock, slice the relevant code, and prompt the LLM to generate inputs. As a result, \tool managed to overcome this roadblock. Subsequently, \tool was iteratively involved in reaching deeper and deeper code locations, and finally exposed the bug. Based on our evaluation, traditional concolic execution engines are incapable of exposing this bug. Condition \ccc{4} is related to a one-way function, which is unsolvable for traditional constraint solvers. Moreover, satisfying condition \ccc{3} requires precise reasoning over code \ccc{2}, which poses another significant challenge for traditional concolic execution.

\result{\tool can be effectively extended to test programs implemented in a different programming language.}





\subsection{Discussion}
We evaluated \tool on widely-used real-world subjects with diverse input types and large codebases. Our results demonstrate that this approach is highly promising, achieving substantial improvements in code coverage compared to the baselines. Currently, our findings are limited to text-based inputs, including both string-based and arithmetic-based inputs. However, these findings may not be generalized to other types of binary-based inputs, as the current LLMs face challenges in generating non-text responses. Furthermore, the applicability of \tool is currently restricted to open-source programs, as it relies on code slices extracted from the source code.

\subsection{Limitations}
Existing hybrid fuzzers execute locally on CPU-based hardware, whereas \tool relies on a remotely hosted LLM service with asymmetric computational resources (e.g., GPUs).
As a result, the reported wall-clock performance also reflects differences in the underlying execution environments.
However, our primary goal is to evaluate the practical effectiveness of LLM-assisted hybrid fuzzing as an end-to-end workflow, rather than hardware-normalized solver performance.

Another limitation concerns the possibility of data leakage in LLM-based systems.
Since modern LLMs are trained on massive corpora of public data, it is possible that some or all of the evaluated programs appeared in the training data, which may influence the LLM's ability to reason about program behavior.
At the same time, \tool also demonstrated effectiveness on the AIxCC Jenkins case study, which was released {\em after} the \verb+gpt-4o+ snapshot used in our experiments.
That said, evaluating the extent of this influence remains an open research question.

\section{Related Work} \label{sec:related_work}

\subsection{{Hybrid Fuzzing}}

Hybrid fuzzing was first introduced through hybrid concolic testing \cite{rapuk2007hybrid}, which combines random testing with concolic execution. \driller \cite{driller} was the first tool to integrate greybox fuzzing with concolic execution, demonstrating effectiveness in uncovering vulnerabilities in the 2016 DARPA's Cyber Grand Challenge subjects. However, \driller struggles to scale to large real-world programs due to the performance bottlenecks inherent in concolic execution. To address this limitation, \qsym \cite{qsym} introduces fast symbolic emulation, accelerated constraint solving, and efficient environment modeling. Building on this, \intriguer \cite{intriguer} further improves performance by adopting field-level constraint solving. Continuing these works, \tool also aims to improve the efficacy of concolic execution, but by leveraging recent advancements in large language models (LLMs) and eliminating the need for translation from a high-level programming language to logic formulas. 

In contrast to improving the performance of concolic execution, several hybrid fuzzers \cite{digfuzz, meuzz, pangolin, hybrid} aim to improve seed synchronization and scheduling. \digfuzz \cite{digfuzz} employs a Monte Carlo-based probabilistic path prioritization model to guide concolic execution to solve the most challenging paths. \meuzz \cite{meuzz} leverages machine learning to predict which seeds are likely to yield better fuzzing results, while \pangolin \cite{pangolin} preserves exploration states to guide mutation and constraint solving. Based on evaluating all previous hybrid fuzzers, \cofuzz \cite{hybrid}  introduced a coordinated strategy to better integrate greybox fuzzing and concolic execution. These general strategies are applicable to various hybrid fuzzers, including \tool, to enhance overall performance. However, \tool itself does not target improving seed synchronization and scheduling.

\subsection{{Concolic Execution}}

\dart \cite{dart} is the first concolic testing tool, pioneering the combination of concrete and symbolic execution to incrementally generate test inputs. \cute \cite{cute} extends \dart to handle multithreaded programs. \exe \cite{exe} performs mixed concrete and symbolic execution, while \klee \cite{klee} is a redesign of \exe that can check a large number of different software systems. \ste \cite{s2e} applies concolic execution to binary code. These foundational works lay the groundwork for concolic execution, from the core concept to real-world use. \tool follows the principles of traditional concolic execution, but we explore the potential to develop smart concolic execution, with knowledge of the program semantics during input generation. 

With the increase of real-world programs in size and complexity, concolic execution shows limitations in checking large programs, such as significant time overhead from symbolic emulation and constraint solving, as well as the manual effort required in environment modeling. To scale to large real-world programs, several optimizations are proposed to address these limitations. \qsym \cite{qsym} uses concrete execution to model external environments and collects an incomplete set of constraints to improve efficiency. \intriguer \cite{intriguer} uses solvers for complicated constraints only and lowers the priority of instructions that use a wide range of input offsets. While these optimizations reduce time overhead and modeling efforts, they may also compromise the performance of concolic execution. Our tool seeks to address these limitations at their core by leveraging the knowledge of the LLMs. The recent study \cite{jiang2024towards} shows that LLMs can reason over constraints comparable to traditional tools such as \klee in toy programs and even outperform them in programs with implicit data flow or out-of-code constraints. These findings further validate the promise of our approach.

\subsection{{LLM-based Unit Test Generation}}

Recent studies \cite{schafer2023empirical, yang2024evaluation} have investigated the capabilities of LLMs in assisting in automatic unit test generation. Existing approaches use the LLMs to generate unit tests for programs in different programming languages. For example, \testpilot \cite{testpilot} prompts LLMs to generate JavaScript unit tests, while \chatunittest \cite{chatunitest} and \hits \cite{HiTS} focus on Java unit tests. Similarly, \codamosa \cite{codamosa}, \coverup \cite{coverup}, and \symprompt \cite{symprompt} guide the LLM to generate unit tests for Python programs. These prompts include natural language descriptions of the intended code functionality, the expected properties, and the source code of the methods under test. These works primarily aim to improve the quality and effectiveness of unit tests, while \tool focuses on improving system testing through fuzzing.

\tool also shares some similarities with these works. Like \codamosa and \chatunittest, \tool incorporates source code into LLM prompts. However, rather than including entire modules, \tool extracts only the relevant path constraints for a specific execution path using program analysis, and then guides the LLM to generate new inputs by reasoning over these constraints. In contrast, \codamosa and \chatunittest rely on full source code, which can hinder coverage when methods involve complex or numerous path conditions. This is a limitation noted by prior work such as \hits \cite{HiTS}. To mitigate this, \hits proposes decomposing focal methods into smaller slices and generating tests for each slice. However, these slices are LLM-generated fragments that may span multiple or partial execution paths and are not grounded in formal static or dynamic slicing techniques. Moreover, \hits incurs additional LLM costs for code slicing, reportedly exceeding two dollars per large file. In comparison, \tool employs dynamic slicing to extract path constraints aligned with specific execution goals, thereby avoiding unnecessary LLM overhead and eliminating additional slicing costs. 

\subsection{{LLM-based Fuzzing}}

LLMs have also been employed in various fuzzers to enhance test generation during fuzz testing. \titanfuzz \cite{titanfuzz} uses the LLMs to generate test cases for deep learning software libraries. \chatafl \cite{chatafl} uses LLMs to enhance network fuzzing. \llmfuzz \cite{llm4fuzz} uses the LLMs to improve test case generation while fuzzing smart contracts. \fuzzall \cite{fuzz4all} advances compiler testing by leveraging LLMs. In these fuzzers, the interleaving between LLMs and greybox fuzzing is similar to that of hybrid fuzzers. The LLMs are usually queried only when greybox fuzzing reaches a coverage plateau, due to the expensive cost of the LLM calls. Similarly, hybrid fuzzers, including \tool, invoke the heavy concolic execution whenever facing a fuzzing plateau. 

However, unlike existing fuzzers, \tool assists greybox fuzzing in a fundamentally different way. Existing approaches exploit the extensive domain knowledge of LLMs---such as knowledge of deep-learning libraries, network protocols, and user specifications---to generate valid inputs during fuzzing. For example, \chatafl generates new inputs based on the LLM's understanding of network protocols, including input formats and state machines. These approaches do not dive into the codebases or reason over code paths. In contrast, our fuzzer \tool uses LLMs to generate new inputs by its reasoning over the code path constraints in the form of code slices, and the new inputs aim to satisfy these path constraints. This makes \tool analogous to traditional concolic execution integrated into hybrid fuzzing, while \tool enhances traditional concolic execution through LLM-based reasoning. To the best of our knowledge, no previous works leverage LLMs to improve the efficacy of hybrid fuzzing.

\subsection{LLM-based Path Reasoning}

Recent studies \cite{llmdfa, llift} have also leveraged LLMs to reason about path constraints, though in different contexts and with different research goals from \tool. \llmdfa focuses on analyzing dataflow information, where LLMs are used to extract source and sink locations, and then summarize their dataflows based on path constraints. On the other hand, \llift aims to improve the performance of detecting Use-Before-Initialization bugs; here, LLMs are used to identify variable initializers and their post conditions, and subsequently determine whether variables are properly initialized. Both \llmdfa and \llift are designed to advance static analysis. In contrast, \tool leverages LLMs to reason over path constraints to enable smart concolic execution, thereby improving the fuzzing performance. 

\subsection{LLM-based Symbolic Execution}

A closely related line of work is {\em LLM-based symbolic execution}, as implemented by {\sc AutoBug}~\cite{li25autobug}, which uses LLMs for path-based program analysis.
While \tool and {\sc AutoBug} share the common idea of leveraging LLMs to reason about program behavior, they represent two distinct design points in the space of LLM-assisted program analysis.
The key differences are summarized in \autoref{tab:hyllfuzz-autobug}.

\begin{table}[t]
\centering
\caption{\tool and {\sc AutoBug}~\cite{li25autobug} comparison.}
\label{tab:hyllfuzz-autobug}
\begin{tabular}{lcc}
\toprule
\textbf{Aspect} & \tool & \textsc{AutoBug} \\
\midrule
{\em Execution model} & Concolic & Symbolic \\
{\em Goal} & Fuzzing & Analysis / verification \\
{\em Constraints} & Dynamic slicing & Symbolic slicing \\
{\em Guarantees} & None (fuzzing) & Formal foundation \\
{\em Coverage} & Edge & Path \\
{\em Cost} & Lower & Higher \\
\bottomrule
\end{tabular}
\end{table}

Conceptually, \tool optimizes for \emph{practical input-space exploration} under fuzzing, whereas {\sc AutoBug} targets \emph{systematic path reasoning} for program analysis.
As such, \tool emphasizes {\em efficiency} and {\em scalability} in testing scenarios, whereas {\sc AutoBug} emphasizes principled reasoning over program paths.
To highlight the differences, we consider the loop in lines 7--17 from \autoref{fig:bkg_code}.
{\em AutoBug} systematically explores path summaries via path decomposition.
In contrast, \tool only considers the subset of paths {\em driven by coverage-guided fuzzing}, and will invoke LLM-based reasoning only when a roadblock is encountered.

Considering only the loop, {\sc AutoBug} explores up to {\bf 7} distinct path summaries derived from the code structure, each potentially triggering an LLM invocation.
In contrast, \tool explores at most {\bf 3} paths to achieve edge coverage in this example, and invokes the LLM only if a roadblock is encountered.
If the other 20+ conditionals are considered, the number of unique path summaries can increase combinatorially.
This difference reflects a fundamental trade-off: \tool reduces analysis cost by focusing on \emph{reachable and coverage-relevant paths}, whereas {\sc AutoBug} aims for systematic path exploration.
\tool is therefore more suitable for large-scale fuzzing scenarios where exhaustive path reasoning is infeasible.

\section{Conclusion} \label{sec:conclusion}

In this paper, we present \tool, a hybrid fuzzer that combines greybox fuzzing with smart concolic execution powered by large language models (LLMs). Taking advantage of recent advances in LLMs, \tool explores the potential to develop smart concolic execution, with knowledge of the code semantics during input generation. In our evaluation, \tool shows strong promise in terms of effectiveness, efficiency, and usability. Our evaluation results show that \tool can cover 31\%--59\% more code branches than the state-of-the-art hybrid fuzzers \qsym, \intriguer, and \cofuzz, leading to more bugs found. In addition, \tool reduces the time usage of each concolic execution run by 3--19 times. Moreover, \tool avoids the combination of a heavy symbolic execution infrastructure and reduces the manual effort involved in environment modeling and system call support, which naturally enhances the usability of hybrid fuzzing and lowers the barrier to its adoption. 



\bibliographystyle{IEEEtran}
\bibliography{reference}

\end{document}